\documentclass[%
reprint,
superscriptaddress,
showpacs,preprintnumbers,
 amsmath,amssymb,
 aps,
 prc,
]{revtex4-1}

\usepackage{graphicx}
\usepackage{dcolumn}
\usepackage{bm}

\begin{document}
\title{\boldmath Missing-mass spectroscopy of the ${}^{12}{\rm C}(p,d)$ reaction  \\ near the $\eta^\prime$-meson production threshold}
\author{Y.~K.~Tanaka}
\email{E-mail: y.tanaka@gsi.de}
\altaffiliation[Present address: ]{GSI Helmholtzzentrum f\"ur Schwerionenforschung GmbH, Planckstra\ss e 1, 64291 Darmstadt, Germany}
\affiliation{The University of Tokyo, 7-3-1 Hongo, Bunkyo, 113-0033 Tokyo, Japan}
\author{K.~Itahashi}
\affiliation{Nishina Center for Accelerator-Based Science, RIKEN, 2-1 Hirosawa, Wako, 351-0198 Saitama, Japan}
\author{H.~Fujioka}
\affiliation{Kyoto University, Kitashirakawa-Oiwakecho, Sakyo-ku, 606-8502 Kyoto, Japan}
\author{Y.~Ayyad}
\affiliation{RCNP, Osaka University, 10-1 Mihogaoka, Ibaraki, 567-0047 Osaka, Japan}
\author{J.~Benlliure}
\affiliation{Universidade de Santiago de Compostela, 15782 Santiago de Compostela, Spain}
\author{K.-T.~Brinkmann}
\affiliation{Universit\"{a}t Giessen, Heinrich-Buff-Ring 16, 35392 Giessen, Germany}
\author{S.~Friedrich}
\affiliation{Universit\"{a}t Giessen, Heinrich-Buff-Ring 16, 35392 Giessen, Germany}
\author{H.~Geissel}
\affiliation{Universit\"{a}t Giessen, Heinrich-Buff-Ring 16, 35392 Giessen, Germany}
\affiliation{GSI Helmholtzzentrum f\"ur Schwerionenforschung GmbH, Planckstra\ss e 1, 64291 Darmstadt, Germany}
\author{J.~Gellanki}
\affiliation{KVI-CART, University of Groningen, Zernikelaan 25, 9747 AA Groningen, the Netherlands}
\author{C.~Guo}
\affiliation{Beihang University, Xueyuan Road 37, Haidian District, 100191 Beijing, China}
\author{E.~Gutz}
\affiliation{Universit\"{a}t Giessen, Heinrich-Buff-Ring 16, 35392 Giessen, Germany}
\author{E.~Haettner}
\affiliation{GSI Helmholtzzentrum f\"ur Schwerionenforschung GmbH, Planckstra\ss e 1, 64291 Darmstadt, Germany}
\author{M.~N.~Harakeh}
\affiliation{KVI-CART, University of Groningen, Zernikelaan 25, 9747 AA Groningen, the Netherlands}
\author{R.~S.~Hayano}
\affiliation{The University of Tokyo, 7-3-1 Hongo, Bunkyo, 113-0033 Tokyo, Japan}
\author{Y.~Higashi}
\affiliation{Nara Women's University, Kita-Uoya Nishi-Machi, 630-8506 Nara, Japan}
\author{S.~Hirenzaki}
\affiliation{Nara Women's University, Kita-Uoya Nishi-Machi, 630-8506 Nara, Japan}
\author{C.~Hornung}
\affiliation{Universit\"{a}t Giessen, Heinrich-Buff-Ring 16, 35392 Giessen, Germany}
\author{Y.~Igarashi}
\affiliation{KEK, 1-1 Oho, Tsukuba, 305-0801 Ibaraki, Japan}
\author{N.~Ikeno}
\affiliation{Tottori University, 4-101 Koyamacho-minami, 680-8551 Tottori, Japan}
\author{M.~Iwasaki}
\affiliation{Nishina Center for Accelerator-Based Science, RIKEN, 2-1 Hirosawa, Wako, 351-0198 Saitama, Japan}
\author{D.~Jido}
\affiliation{Tokyo Metropolitan University, 1-1 Minami-Osawa, Hachioji, 192-0397 Tokyo, Japan}
\author{N.~Kalantar-Nayestanaki}
\affiliation{KVI-CART, University of Groningen, Zernikelaan 25, 9747 AA Groningen, the Netherlands}
\author{R.~Kanungo}
\affiliation{Saint Mary's University, 923 Robie Street, Halifax, Nova Scotia B3H 3C3, Canada}
\author{R.~Knoebel}
\affiliation{Universit\"{a}t Giessen, Heinrich-Buff-Ring 16, 35392 Giessen, Germany}
\affiliation{GSI Helmholtzzentrum f\"ur Schwerionenforschung GmbH, Planckstra\ss e 1, 64291 Darmstadt, Germany}
\author{N.~Kurz}
\affiliation{GSI Helmholtzzentrum f\"ur Schwerionenforschung GmbH, Planckstra\ss e 1, 64291 Darmstadt, Germany}
\author{V.~Metag}
\affiliation{Universit\"{a}t Giessen, Heinrich-Buff-Ring 16, 35392 Giessen, Germany}
\author{I.~Mukha}
\affiliation{GSI Helmholtzzentrum f\"ur Schwerionenforschung GmbH, Planckstra\ss e 1, 64291 Darmstadt, Germany}
\author{T.~Nagae}
\affiliation{Kyoto University, Kitashirakawa-Oiwakecho, Sakyo-ku, 606-8502 Kyoto, Japan}
\author{H.~Nagahiro}
\affiliation{Nara Women's University, Kita-Uoya Nishi-Machi, 630-8506 Nara, Japan}
\author{M.~Nanova}
\affiliation{Universit\"{a}t Giessen, Heinrich-Buff-Ring 16, 35392 Giessen, Germany}
\author{T.~Nishi}
\affiliation{Nishina Center for Accelerator-Based Science, RIKEN, 2-1 Hirosawa, Wako, 351-0198 Saitama, Japan}
\author{H.~J.~Ong}
\affiliation{RCNP, Osaka University, 10-1 Mihogaoka, Ibaraki, 567-0047 Osaka, Japan}
\author{S.~Pietri}
\affiliation{GSI Helmholtzzentrum f\"ur Schwerionenforschung GmbH, Planckstra\ss e 1, 64291 Darmstadt, Germany}
\author{A.~Prochazka}
\affiliation{GSI Helmholtzzentrum f\"ur Schwerionenforschung GmbH, Planckstra\ss e 1, 64291 Darmstadt, Germany}
\author{C.~Rappold}
\affiliation{GSI Helmholtzzentrum f\"ur Schwerionenforschung GmbH, Planckstra\ss e 1, 64291 Darmstadt, Germany}
\author{M.~P.~Reiter}
\affiliation{GSI Helmholtzzentrum f\"ur Schwerionenforschung GmbH, Planckstra\ss e 1, 64291 Darmstadt, Germany}
\author{J.~L.~Rodr\'{i}guez-S\'{a}nchez}
\affiliation{Universidade de Santiago de Compostela, 15782 Santiago de Compostela, Spain}
\author{C.~Scheidenberger}
\affiliation{Universit\"{a}t Giessen, Heinrich-Buff-Ring 16, 35392 Giessen, Germany}
\affiliation{GSI Helmholtzzentrum f\"ur Schwerionenforschung GmbH, Planckstra\ss e 1, 64291 Darmstadt, Germany}
\author{H.~Simon}
\affiliation{GSI Helmholtzzentrum f\"ur Schwerionenforschung GmbH, Planckstra\ss e 1, 64291 Darmstadt, Germany}
\author{B.~Sitar} 
\affiliation{Comenius University Bratislava, Mlynsk\'{a} dolina, 842 48 Bratislava, Slovakia}
\author{P.~Strmen}
\affiliation{Comenius University Bratislava, Mlynsk\'{a} dolina, 842 48 Bratislava, Slovakia}
\author{B.~Sun}
\affiliation{Beihang University, Xueyuan Road 37, Haidian District, 100191 Beijing, China}
\author{K.~Suzuki}
\affiliation{Stefan-Meyer-Institut f\"{u}r subatomare Physik, Boltzmangasse 3, 1090 Vienna, Austria}
\author{I.~Szarka}
\affiliation{Comenius University Bratislava, Mlynsk\'{a} dolina, 842 48 Bratislava, Slovakia}
\author{M.~Takechi}
\affiliation{Niigata University, 8050 Ikarashi 2-no-cho, Nishi-ku, 950-2181 Niigata, Japan}
\author{I.~Tanihata} 
\affiliation{RCNP, Osaka University, 10-1 Mihogaoka, Ibaraki, 567-0047 Osaka, Japan}
\affiliation{Beihang University, Xueyuan Road 37, Haidian District, 100191 Beijing, China}
\author{S.~Terashima}
\affiliation{Beihang University, Xueyuan Road 37, Haidian District, 100191 Beijing, China}
\author{Y.~N.~Watanabe}
\affiliation{The University of Tokyo, 7-3-1 Hongo, Bunkyo, 113-0033 Tokyo, Japan}
\author{H.~Weick}
\affiliation{GSI Helmholtzzentrum f\"ur Schwerionenforschung GmbH, Planckstra\ss e 1, 64291 Darmstadt, Germany}
\author{E.~Widmann}
\affiliation{Stefan-Meyer-Institut f\"{u}r subatomare Physik, Boltzmangasse 3, 1090 Vienna, Austria}
\author{J.~S.~Winfield}
\affiliation{GSI Helmholtzzentrum f\"ur Schwerionenforschung GmbH, Planckstra\ss e 1, 64291 Darmstadt, Germany}
\author{X.~Xu}
\affiliation{GSI Helmholtzzentrum f\"ur Schwerionenforschung GmbH, Planckstra\ss e 1, 64291 Darmstadt, Germany}
\author{H.~Yamakami}
\affiliation{Kyoto University, Kitashirakawa-Oiwakecho, Sakyo-ku, 606-8502 Kyoto, Japan}
\author{J.~Zhao}
\affiliation{Beihang University, Xueyuan Road 37, Haidian District, 100191 Beijing, China}

\collaboration{$\eta$-PRiME/Super-FRS Collaboration}

\date{\today}

\begin{abstract}

Excitation-energy spectra of $^{11}$C nuclei
near the $\eta^\prime$-meson 
production threshold
have been measured by missing-mass spectroscopy
using the $^{12}$C($p$,$d$) reaction.
A carbon target has been irradiated with a 2.5 GeV proton beam supplied by
the synchrotron SIS-18 at GSI to produce $\eta^\prime$ meson bound states
in $^{11}$C nuclei. 
Deuterons emitted at $0^\circ$ in the reaction have been momentum-analyzed 
by the fragment separator (FRS) used as a high-resolution spectrometer. 
No distinct structure due to the formation of $\eta^\prime$-mesic states
is observed  
although a high statistical sensitivity is achieved 
in the experimental spectra.  
Upper limits on the formation cross sections 
of $\eta^\prime$-mesic states are determined,
and thereby a constraint imposed on the $\eta^\prime$-nucleus 
interaction is discussed.

\end{abstract}


\maketitle 
\section{Introduction}

Understanding 
hadron masses is one of the major subjects
in contemporary hadron physics. 
Studies of the light pseudoscalar mesons are of particular importance, 
since they have close relations 
to fundamental symmetries in quantum chromodynamics (QCD).
The flavor-octet mesons are considered to be
Nambu-Goldstone bosons associated with spontaneous breaking of chiral symmetry, 
leading to relatively small masses of the $\pi$, $K$, and $\eta$ mesons.
In contrast, the  $\eta^\prime$ meson 
has an exceptionally large mass of $958$~MeV/$c^2$, 
which has attracted interest known as the ``U(1) problem'' \cite{Weinberg1975}. 
Theoretically, the large $\eta^\prime$ mass
can be explained by the explicit breaking of $U_A(1)$ symmetry 
owing to quantum anomaly effects in QCD \cite{Witten1979, Veneziano1979}. 
This anomaly effect on the $\eta^\prime$ mass 
is expected to manifest itself under  
the presence of chiral symmetry breaking \cite{Lee1996, Jido2012}.

In finite baryon density, properties of the mesons 
may be modified from those in the vacuum due to 
partial restoration of chiral symmetry \cite{Hayano2010, Leupold2010, Metag2017}.
For the $\eta^\prime$ meson, the reduction of the mass is expected 
through a weakening of the anomaly effect \cite{Jido2012}
and also predicted in various theoretical models.
For example,  
about 150~MeV/$c^2$ reduction at the nuclear saturation density
is predicted by the Nambu--Jona-Lasinio (NJL) model \cite{Costa2005,Nagahiro2006}, 
80~MeV/$c^2$ by the linear sigma model \cite{Sakai2013}, 
and 37~MeV/$c^2$ by the quark meson coupling (QMC) model \cite{Bass2006}.
Investigation of such a modification would yield novel insights 
into the mechanism of the meson mass generation as well 
as the vacuum structure of QCD.

Meson-nucleus bound states open a unique possibility of 
directly probing in-medium meson properties.
A well-established example of such systems is  
the deeply-bound $\pi^{-}$ states in heavy nuclei, where 
a $\pi^{-}$ meson is bound near the nuclear surface 
by the superposition of the attractive Coulomb interaction
and the repulsive $s$-wave strong interaction.
These states have been discovered and studied 
in missing-mass spectroscopy of the ($d$,$^3$He) reaction 
\cite{Yamazaki1996, ItahashiGilg2000, Geissel2002}.
A large overlap between a $\pi^{-}$ meson and a nucleus 
in well-defined quantum bound states
allows the extraction of  
information on a modification 
of the isovector part of the $s$-wave pion-nucleus potential,
leading to a quantitative evaluation of partial 
restoration of chiral symmetry at finite nuclear density \cite{Suzuki2004, Kolomeitsev2003, Jido2008, Yamazaki2012}.

In the case of neutral mesons \cite{Metag2017}, 
bound states may be formed only via the strong interaction,
if the attraction between a meson and a nucleus is strong enough.
In-medium meson properties, 
the mass shift $\Delta m (\rho_0)$ and width $\Gamma (\rho_0)$ at the nuclear saturation density $\rho_0$,  
are incorporated
in the meson-nucleus potential as
$U(r)=(V_0 + i W_0) \rho (r)/\rho_0$ 
by relations $V_0=\Delta m(\rho_0)$ and $W_0=$ $-\Gamma(\rho_0)/2$,
where $\rho(r)$ denotes the nuclear density distribution.
A small imaginary potential compared with the real part,
i.e., $|W_0|<|V_0|$, is required for 
the existence of bound states as discrete levels.
This condition may be satisfied for the $\eta^\prime$ meson, 
as described below.

Very limited information 
on the $\eta^\prime$-nucleus interaction is currently available.
On the theoretical side, the predictions 
for the $\eta^\prime$-mass 
reduction \cite{Costa2005,Nagahiro2006,Sakai2013,Bass2006}
suggest the real part of the potential 
$V_0$ in the range from $-150$~MeV to $-37$~MeV.
On the experimental side, 
the CBELSA/TAPS collaboration deduced
the real part 
as $V_0 = -(39 \pm 7(\mathrm{stat}) \pm 15(\mathrm{syst}))$~MeV 
from $\eta^\prime$ momentum distributions and 
excitation functions in  
$\eta^\prime$ photo-production on nuclear targets \cite{Nanova2013,Nanova2016}.
They also evaluated the imaginary part of $W_0 = -(13 \pm 3(\mathrm{stat}) \pm 3(\mathrm{syst}))$~MeV 
by measuring transparency ratios 
as a function of the mass number of the target nuclei \cite{Nanova2012} 
and as a function of $\eta^\prime$ momentum \cite{Friedrich2016}.
Such a small imaginary part relative to the real part 
implies the  
possibility of observing a bound state as a distinct peak structure. 
In the meantime, the real part of the scattering length for the $\eta^\prime$-proton interaction 
has been
extracted from measurements of the $pp\rightarrow pp\eta^\prime$ reaction close to its threshold 
to be 
$0.00 \pm 0.43$~fm \cite{Czerwinski2014}, 
corresponding to an $\eta^\prime$-nucleus potential depth of $|V_0|<38$~MeV 
at the nuclear density of 0.17~fm$^{-3}$ within the low-density approximation. 

Experimental programs to search for $\eta^\prime$-mesic nuclei 
have recently been started, aiming 
at directly studying the in-medium properties of the $\eta^\prime$ meson.
A one-nucleon pickup reaction, 
for example ($p, d$) or ($\gamma, p$),
at forward angles is preferable to produce the $\eta^\prime$-nucleus 
bound states,
because the momentum transfer of such a reaction can be rather small.
Spectroscopy experiments of the $^{12}$C($\gamma, p$) reaction \cite{Nagahiro2005}
using high-energy photon beams were proposed 
by the LEPS2 collaboration at the SPring-8 facility \cite{Muramatsu_Baryon2013} 
and by the BGO-OD collaboration at the ELSA accelerator \cite{Metag_BGOOD_Proposal}.
The results of these experiments are thus far not available.

We proposed an experimental search for $\eta^\prime$-mesic nuclei  
with missing-mass spectroscopy of the $^{12}$C($p$,$d$) reaction 
near the $\eta^\prime$ production threshold 
\cite{Itahashi2012}.
The kinetic energy of the proton beam was 
chosen to be 2.5~GeV, slightly above the threshold energy 
for the elementary process 
$n(p,d)\eta^\prime$
of 2.4~GeV.
The momentum transfer of this reaction at 0$^\circ$
is moderate ($\sim 500$~MeV/$c$) at 2.5~GeV.
An inclusive measurement of the forward-emitted deuterons allows 
the analysis of 
the overall ($p$,$d$) spectrum without any assumption on
decay processes of $\eta^\prime$-mesic nuclei.

The formation cross section of the $\eta^\prime$-mesic nuclei 
via the $^{12}$C($p, d$)$^{11}$C$\otimes \eta^\prime$ 
reaction has been theoretically calculated in Ref.~\cite{Nagahiro2013} 
for various sets of $(V_0, W_0)$, the real and imaginary parts  
of the $\eta^\prime$-nucleus potential.
Population of $\eta^\prime$-mesic states coupling 
with neutron hole states has been predicted,
depending on the assumed potential.
Distinct peak structures are expected in the excitation spectra 
particularly near the $\eta^\prime$ production threshold
because of the enhanced excited states due to the finite momentum transfer of the reaction \cite{Nagahiro2013}. 

Physical background such as quasi-free meson production,
$pN \rightarrow dX$ ($X=2\pi, 3\pi, 4\pi, \omega$), 
also contributes to the experimental spectrum as a continuum.
The cross section of the above background process was estimated to be  
2--3 orders of magnitude larger than 
that of the formation of the $\eta^\prime$-mesic states \cite{Itahashi2012}.
To overcome such a small signal-to-background ratio, 
we aimed at achieving an extremely high statistical 
sensitivity with relative errors of $< 1$\% in the spectrum. 
An inclusive simulation 
has shown that observing peak structures 
near the threshold
is feasible under a realistic experimental 
condition for
a strongly attractive potential $V_0 \lesssim -100$~MeV \cite{Itahashi2012}.

We carried out the 
experiment in 2014.  
While its 
results  
have been briefly reported 
elsewhere \cite{Tanaka2016},
in this paper a full description of the experiment and analysis
including discussions based on additional theoretical calculations is presented.
First, the experimental method and performed measurements are introduced (Sec.~\ref{s_experiment}), 
and next the data analysis to obtain the excitation-energy spectra 
is described in detail (Sec.~\ref{s_analysis}). 
The results of a statistical analysis for the obtained spectra are explained 
(Sec.~\ref{s_results}),
followed by discussions of the $\eta^\prime$-nucleus interaction and 
future plans for a follow-up experiment with higher sensitivity
(Sec.~\ref{s_discussion}). 
Finally, a conclusion is given in Sec.~\ref{s_conclusion}. 

\section{Experiment\label{s_experiment}}

A missing-mass spectroscopy experiment 
using the $^{12}$C($p$,$d$) reaction 
was performed
near the $\eta^\prime$ production threshold 
at GSI, Darmstadt, Germany. 
A 2.5~GeV proton beam impinged on 
 a carbon target, 
and the emitted deuterons at $0^\circ$ were momentum-analyzed 
to obtain missing masses in the reaction. 
In addition,  
elastic proton-deuteron scattering was measured 
for 
the calibration of the experimental system.

\subsection{Proton beam \label{ss_beam}}
Proton beams were supplied by the synchrotron SIS-18. 
Two kinetic energies were employed: 
$2499.1 \pm 2.0$ MeV for the measurement of the $^{12}$C($p$,$d$) reaction
and $1621.6 \pm 0.8$~MeV 
for the calibration 
with proton-deuteron elastic scattering. 
These energies were determined by 
measuring precisely the revolution frequencies of the beams in the synchrotron.
The accelerated beams were extracted in a slow extraction mode with a 
spill length of 4 (1) seconds and a cycle of 7 (4) seconds 
for 2.5 (1.6) GeV.
The beams were focused at the experimental target, 
where a typical spot size was $\sim 1$~mm (horizontal) $\times$ $3$~mm (vertical).

The beam intensity was $\sim 10^{10}$/s, 
measured in front of the target by
the SEETRAM detector \cite{Jurado2002} inserted on the beam axis. 
This detector was used only during a short dedicated measurement for 
an 
absolute normalization of the cross sections
to avoid unnecessary material near the beam axis.
Plastic scintillation counters placed off axis around the target
continuously monitored relative changes in the luminosity 
by counting scattered particles from the target.

\subsection{Target \label{ss_target}}
Three targets were mounted on 
a movable ladder at the entrance position of the spectrometer. 
A natural carbon target 
with an areal density of 4115 $\pm$ 1 mg/cm$^2$ was 
used for the measurement of the $^{12}$C($p$,$d$) reaction. 
Deuterated polyethylene (CD$_2$) targets 
with areal densities of 1027 $\pm$ 2 mg/cm$^2$
and 4022 $\pm$ 9 mg/cm$^2$ were 
used for the calibration via 
the proton-deuteron elastic $\mathrm{D}(p,d)p$ reaction. 
These targets had a cylindrical shape with a diameter of 2~cm.

\subsection{Spectrometer and detector system}
\begin{figure}[hbtp]
\centering \includegraphics[width=85.0mm]{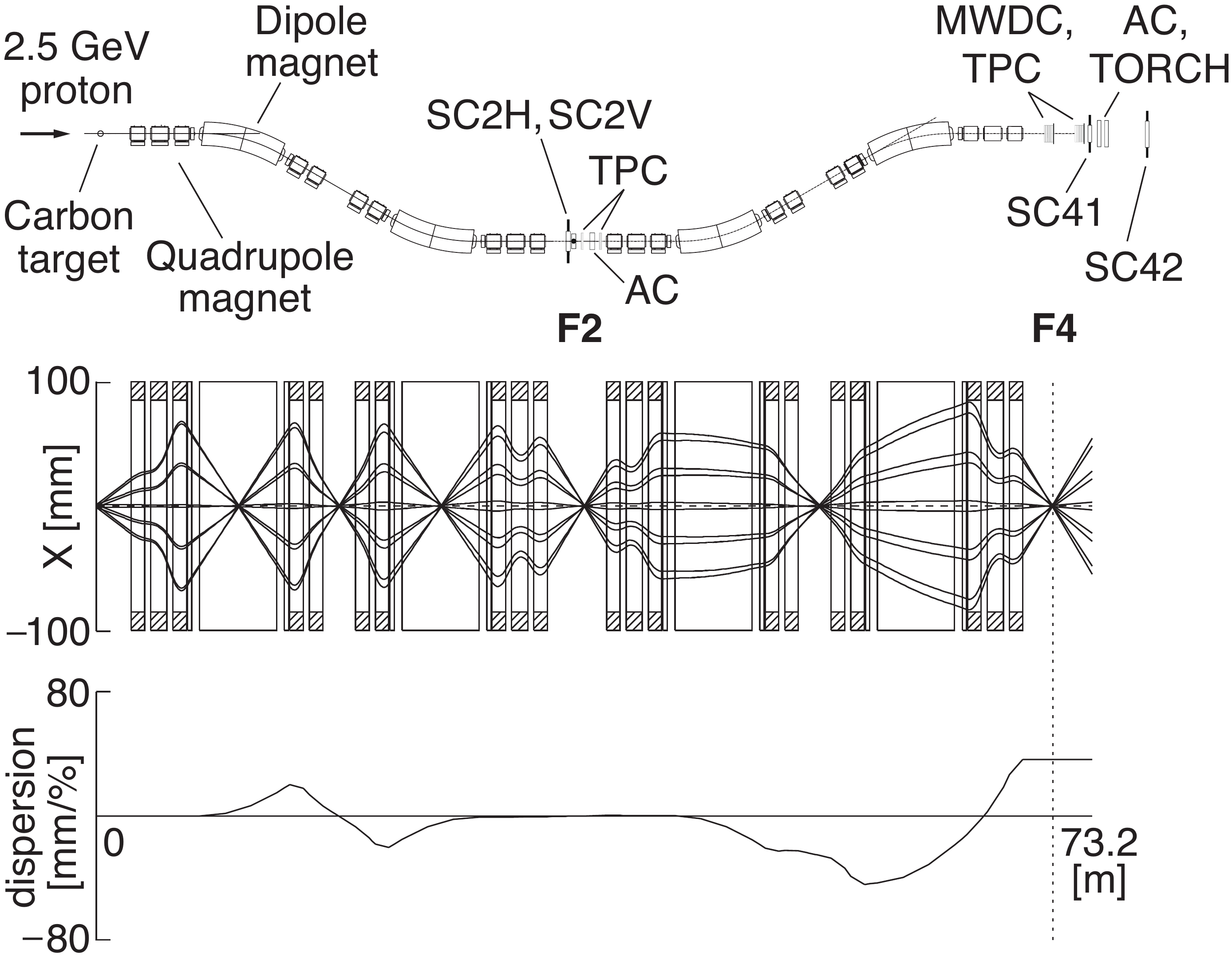}
\caption{\label{fig_optics_setup}
(Top panel) A schematic view of the experimental setup with the FRS.
A 2.5 GeV proton beam impinged on a carbon target. 
Deuterons emitted in the $^{12}$C($p$,$d$) reaction 
were momentum analyzed at F4 and the tracks were 
reconstructed from measurements by multi-wire drift chambers (MWDCs). 
Sets of 5 mm-thick plastic scintillation counters (SC2H, SC2V, and SC41)
and a 20 mm-thick one (SC42) were installed at F2 and F4 
for time-of-flight (TOF) measurements.
\v{C}erenkov detectors (ACs and TORCH) and 
time projection chambers (TPCs) were installed 
but not used in the present analysis.
(Middle panel) Horizontal beam trajectories 
with the specially-developed optics mode
based on calculated third-order transfer matrices. 
Initial positions and angles are taken from 
$\{ -1, 1 \}$~mm $\times$ $\{-8, -4, 0, 4, 8\}$~mrad, and a momentum at the central value.
(Bottom panel) A momentum-dispersion curve 
of this optics mode.
}
\end{figure}

The fragment separator (FRS) \cite{Geissel1992} 
was used 
as a high-resolution magnetic spectrometer 
to precisely analyze the momenta of deuterons 
emitted at 0$^\circ$ in the ($p$,$d$) reactions.
The FRS has four stages,  
as schematically depicted in the top panel of Fig.~\ref{fig_optics_setup}.
Each stage consists of a 30$^\circ$-bending dipole magnet, 
quadrupole doublet and triplet magnets. 
Such a configuration provides considerable flexibility 
to realize various ion-optics modes with high momentum resolving powers.

We developed a special ion-optical mode of the FRS,
which is momentum-achromatic at the central focal plane (F2) 
and dispersive at the final focal plane (F4), 
as illustrated in the middle and bottom panels of Fig.~\ref{fig_optics_setup}.
The achromatic section from the target to F2 
was used 
to select particles originating from reactions in the target.
Secondary background produced by the non-interacting primary beam
dumped near the exit of the first dipole magnet was thus rejected. 
The momenta of the deuterons were then analyzed in the dispersive 
section from F2 to F4, 
where a designed momentum resolving power was about $3.8 \times 10^3$.
The dispersion was kept relatively small  
throughout the whole spectrometer 
to have a wide 
momentum acceptance. At F4 the dispersion was 35.1~mm/\%. 

The detection system is depicted in the top panel of Fig.~\ref{fig_optics_setup}.
Two sets of multi-wire drift chambers (MWDCs) were installed at F4
to reconstruct the deuteron tracks and obtain their momenta. 
A MWDC had eight layers of detection planes, 
each consisting of 48 anode wires with a spacing of 5~mm. 
The active area of each layer was 24~cm (horizontal) $\times$ 14~cm (vertical).
The wires in the first four layers were aligned vertically,  
while those in the next two and the last two layers were inclined by $-15^\circ$ and $+15^\circ$, 
respectively. 
The wire positions in the neighboring layers with the same wire angle
were shifted from 
each other by a half length of the spacing.
These MWDCs were operated with a gas mixture of 76\% argon, 
20\% isobutane, and 4\% dimethoxymethane.
The signals from the anode wires were processed by preamplifier-shaper-discriminator chips,
and the resulting timing information was recorded by time-to-digital convertors.

Plastic scintillation counters (SC2H, SC2V, SC41, and SC42) were  
installed for time-of-flight (TOF) measurements 
to distinguish signal 
deuterons with a velocity $\sim 0.84 \, c$ 
from background protons with $\sim 0.95 \, c$. 
SC2H and SC2V had an active area of 6~cm (horizontal) $\times$ 6~cm (vertical) and a thickness of 5~mm,
whereas SC41 had 24~cm (horizontal) $\times$ 6~cm (vertical) and the same thickness.
A plastic scintillator with an area of 50~cm (horizontal) $\times$ 35~cm (vertical)
and a thickness of 2~cm was adopted for SC42, 
which was located at 5.16~m downstream of SC41.
Each of these scintillators was equipped with two photomultipliers.
The anode signals were recorded by a 1-GHz sampling digitizer,
and the discriminated timing information by a time-to-digital convertor.

In addition, 
aerogel \v{C}erenkov detectors (ACs) 
and a total-reflection \v{C}erenkov detector (TORCH) 
were installed for confirmation of the particle identification.
The ACs had silica aerogel with a refractive index of 1.17 \cite{Tabata2010} as a radiator,
corresponding to a threshold velocity of $0.85c$. 
TORCH was equipped with an acrylic radiator
with a refractive index of 1.5, hence 
the maximum detection velocity of $\sim 0.89 \, c$ due to 
designed insensitivity to 
totally-reflected photons \cite{YoneyamaTORCH}.
Time projection chambers (TPCs) \cite{Janik2011},  
the standard beam diagnostics devices of the FRS, 
were also placed at F2 and F4 for the purpose of the online beam tuning. 
These \v{C}erenkov detectors and TPCs are not used 
in the offline analysis described in this paper.

\subsection{Trigger condition for data acquisition}

\begin{figure}[hbtp]
\centering \includegraphics[width=85.0mm]{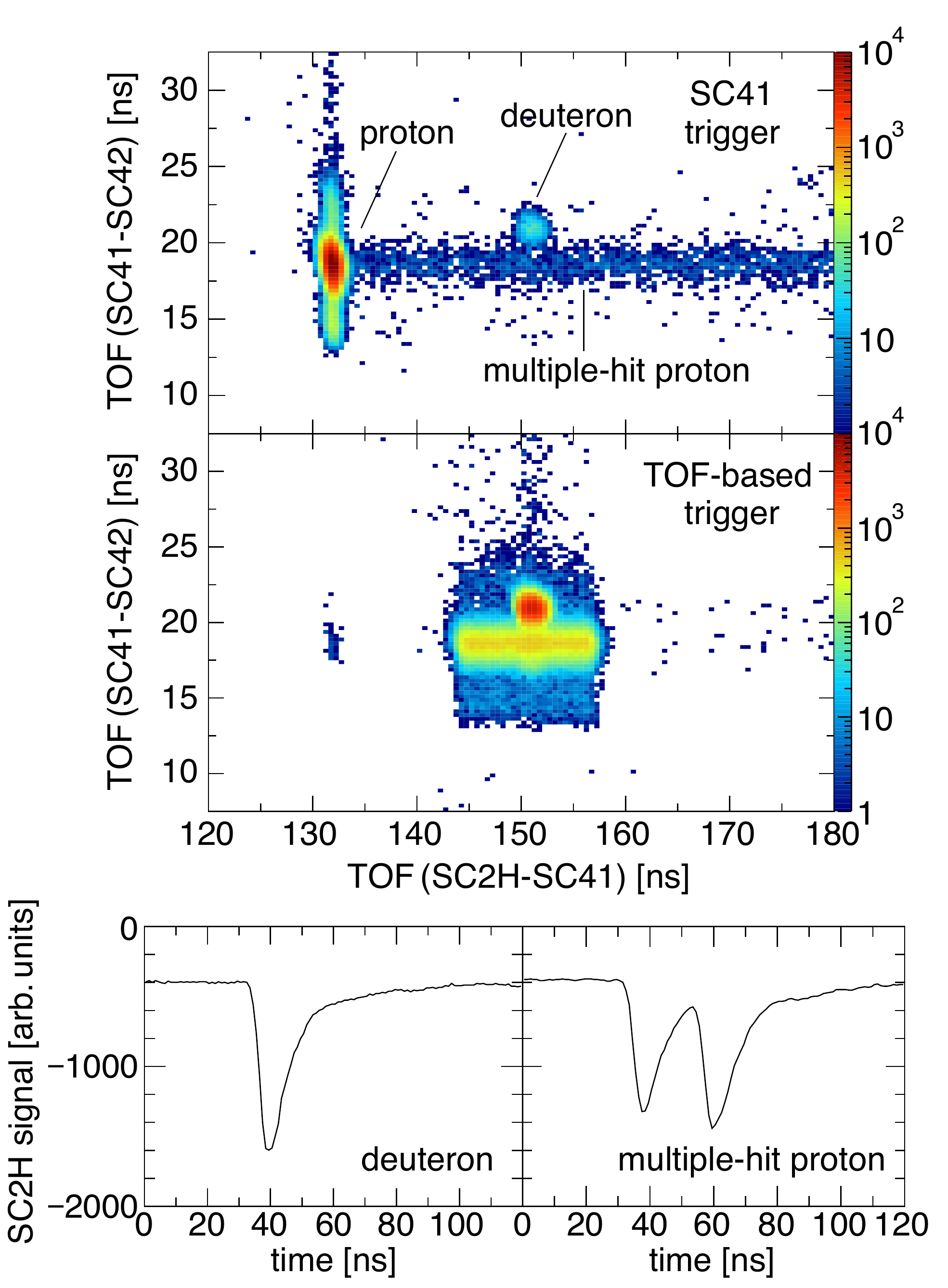}
\caption{\label{fig_tof_online_trigger} 
TOF between SC2H and SC41 and between SC41 and SC42 
recorded with a data acquisition system triggered by the SC41 signals (top) and 
by the TOF-based coincidence between the SC2H and SC41 signals (middle). 
Typical signal waveforms of SC2H are shown
for the deuteron (bottom left) and the multiple-hit proton (bottom right) events based on the TOFs.
}
\end{figure}

In the measurements of the $^{12}$C($p$,$d$) reaction, the total rate of 
charged particles 
at F4 was $\sim 250$~kHz. The top panel 
of Fig.~\ref{fig_tof_online_trigger} shows 
a histogram of TOF between SC2H and SC41 versus one between SC41 and SC42
for all particles reaching F4, obtained with a data-acquisition system 
triggered by the SC41 signal. 
Concentration of events  
corresponding to deuterons and protons
are seen at  
the expected locations and clearly identified. 
The ratio of the number of deuterons to that of protons 
is about 1 to 200, indicating that 
the deuteron and proton rates at F4 were 
$\sim 1$~kHz and 250~kHz, respectively.

In order to reject the background protons at the hardware level,
we employed a TOF-based trigger for the data-acquisition system
by requiring a coincidence of the SC2H and SC41 signals within 15 ns 
around a relative timing of the deuterons.
A TOF histogram with this trigger 
is shown in 
the middle panel of Fig.~\ref{fig_tof_online_trigger}, 
demonstrating efficient rejection of the background protons at the trigger level. 
The main source of the remaining background events was accidental 
multiple-hit protons at SC2H, which are randomly distributed 
in TOF between SC2H and SC41. 
The multiple-hit protons are actually observed in the recorded signal waveforms of SC2H, 
as shown in the bottom panels of Fig.~\ref{fig_tof_online_trigger},
and are rejected later in the data analysis (Sec.~\ref{ss_pid}).
The live rate of the acquisition system varied typically between 30\%--40\%, 
and the data were recorded with a rate of about 10$^3$ events/s. 

\subsection{Summary of measurements}

The experimental conditions are summarized
in Table~\ref{table_measurement_summary}. 
Three types of measurements have been performed, 
which are explained as follows.

\begin{table*}[t]
\caption{\label{table_measurement_summary}
Experimental conditions of the performed measurements.
The first seven rows show production runs of the $^{12}$C($p$,$d$) reaction 
at the proton energy of 2.5 GeV with seven momentum settings of the FRS. 
The next nine rows are calibrations using the elastic D($p$,$d$)$p$ 
reactions at 1.6 GeV with nine FRS settings.
One set of the calibration took about 0.5 hours,
and the number of repetitions is given in the last column.
The last row shows measurements of the elastic D($p$,$d$)$p$ scattering 
at 2.5 GeV. 
}
\begin{ruledtabular}
\begin{tabular}{cccccc}
Reaction & Proton energy & Target               &  Central momentum  & Scaling factor $f$ & Duration  \\
         &    (MeV)      &                      &   of FRS (MeV/$c$) & for magnets &           \\
\hline
 ${}^{12}$C($p$,$d$) & 2499.1 $\pm$ 2.0     &  C (4115 $\pm$ 1 mg/cm$^2$) & 2771.4 & 0.980 & 9.7 hour \\ 
                     &                      &                             & 2779.9 & 0.983 & 9.3 hour \\
                     &                      &                             & 2785.6 & 0.985 & 9.9 hour \\
                     &                      &                             & 2799.7 & 0.990 & 10.9 hour \\ 
                     &                      &                             & 2828.0 & 1.000 & 23.0 hour \\ 
                     &                      &                             & 2856.3 & 1.010 &  5.9 hour \\
                     &                      &                             & 2884.6 & 1.020 &  2.0 hour \\
D($p$,$d$)$p$        & 1621.6 $\pm$ 0.8     &  CD${}_2$ (1027 $\pm$ 2 mg/cm$^2$) 
                                                                          & 2771.4 & 0.980 &  
                                                                          5 set \\
                     &                      &                             & 2779.9 & 0.983 & 
                     2 set \\
                     &                      &                             & 2785.6 & 0.985 &  
                     1 set \\
                     &                      &                             & 2799.7 & 0.990 &  
                     5 set \\                     
                     &                      &                             & 2813.9 & 0.995 &  
                     1 set \\                     
                     &                      &                             & 2828.0 & 1.000 &  
                     8 set \\                     
                     &                      &                             & 2842.1 & 1.005 &  
                     1 set \\                     
                     &                      &                             & 2856.3 & 1.010 &  
                     3 set \\                     
                     &                      &                             & 2884.6 & 1.020 &  
                     1 set \\                     
D($p$,$d$)$p$        & 2499.1 $\pm$ 2.0     &  CD${}_2$ (4022 $\pm$ 9 mg/cm$^2$) & 3809.3 & 1.347 & 1.2 hour
\end{tabular}
\end{ruledtabular}
\end{table*}

The production measurements of the $^{12}$C($p$,$d$) reaction 
were carried out by using the 2.5 GeV proton beam impinging on the  
carbon target.  
The emitted deuterons had a momentum of $2814.4 \pm 2.4$~MeV/$c$  
at the $\eta^\prime$ production threshold
after the energy loss in the target.  
To cover a wide excitation-energy region, 
measurements were conducted at seven central momenta of the FRS
by scaling the whole magnetic field with factors from $f=0.980$ to $f=1.020$.
In particular, a region near the $\eta^\prime$ production threshold was intensively measured,
as distinct narrow structures of $\eta^\prime$-mesic states 
were theoretically predicted  
most strongly near the threshold \cite{Nagahiro2013}.

The momentum calibration of the spectrometer was performed 
by measuring the elastic D($p$,$d$)$p$ reaction at 1.6 GeV using a CD$_2$ target. 
Nearly monochromatic deuterons with the momentum of $2828.0 \pm 1.0$~MeV/$c$ were emitted from the target, 
which defined the central momentum of the FRS at $f=1.000$.
These deuterons were measured with various scale factors 
to analyze the ion-optical response  
of the FRS. 
One calibration run took 
about half an hour, 
and 
it was repeated every $\sim 8$ hours to check 
the stability of the whole spectrometer system.

The elastic D($p$,$d$)$p$ scattering was measured 
with a CD$_2$ target and a proton beam of 2.5~GeV
in order to  
crosscheck the normalization of the differential cross section.
The obtained value was then compared with those reported in Ref.~\cite{Berthet1982}, as explained in Sec.~\ref{ss_normalization}.
During a part of the measurement, 
solid angles were tightly limited to $2.35 \times 10^{-2}$~msr and 
$3.94 \times 10^{-2}$~msr by using slits directly behind the target.
A comparison of the yields with and without the slits 
provided the effective solid angle covered by the FRS.

\section{Analysis \label{s_analysis}}

The goal of the data analysis described in this section 
is to obtain excitation-energy spectra  
of $^{11}$C near the $\eta^\prime$ production threshold. 
The analysis procedure consists of the following steps.
First, the deuteron events are identified (Sec.~\ref{ss_pid}), 
and next the deuteron momenta are reconstructed from the measured tracks (Sec.~\ref{ss_momentum_analysis}).
The excitation energies of $^{11}$C are then kinematically calculated (Sec.~\ref{ss_excitation_energy}), 
and finally the normalization of the cross section is 
performed (Sec.~\ref{ss_normalization}).

\subsection{Selection of deuteron events\label{ss_pid}}

Data of the plastic scintillation counters are analyzed 
in order to identify deuterons at the F4 focal plane. 
As explained in Fig.~2, 
major background particles in the recorded data are the accidental 
multiple-hit protons at F2.
Thus, signal waveforms of photomultipliers reading SC2H 
are firstly analyzed 
to select single-hit events at F2.
TOF between F2 and F4 is analyzed as well  
to further reject remaining background protons. 
SC42 is not used in the following analysis to avoid position-dependent transmission 
from SC41 to SC42 caused by material inhomogeneity found in the 
\v{C}erenkov detectors behind SC41.

The waveforms of SC2H signals are fitted 
by an empirical function
\begin{equation}
f(t) = p_0 + p_1 t - p_2 \cdot \exp{\left( -\frac{(t-p_3)^2}{2(p_4+p_5 t)^2}\right)},\label{f_waveform}
\end{equation}
where the first two terms represent a baseline and the third  
a pulse with a negative polarity.
Signals are fitted within a time window of 70~ns around a typical deuteron time,
treating $p_1, \dots , p_5$ as free parameters.
The sum of the squared residual (SSR) given by the fit is then used to quantify 
the multiplicity of the particles.
Figure~3 shows SSR against the height parameter $p_2$ obtained 
for both two photomultipliers (left and right) of SC2H. 
Single-hit events are clearly identified around SSR $\approx 100$.
By taking into account the correlation between SSR and $p_2$, 
events in the regions indicated by the arrows are selected
as single-hit events and used 
in the following analyses.
 
\begin{figure}[hbtp]
\centering \includegraphics[width=85.0mm]{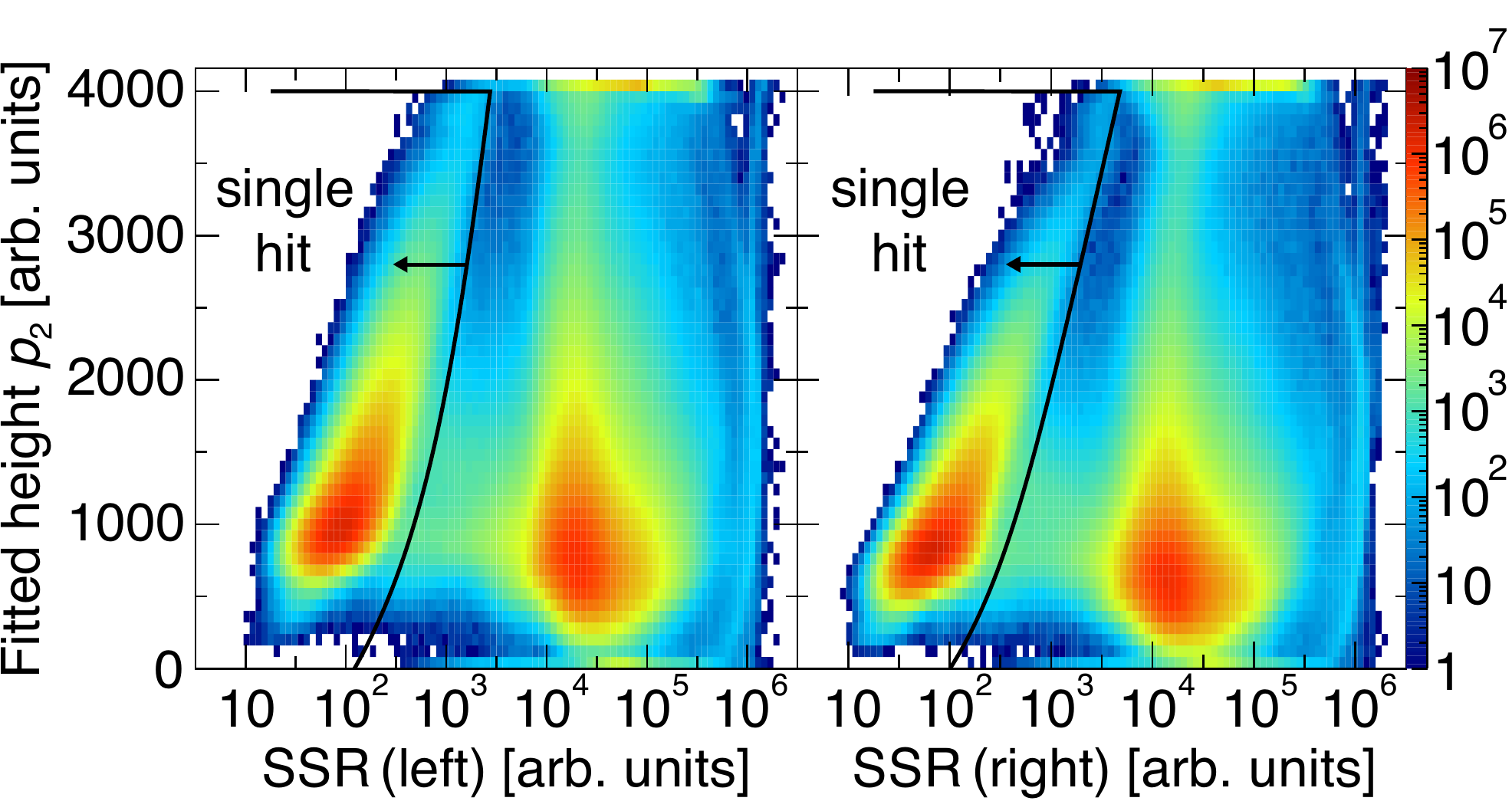} 
\caption{\label{fig_pid_fadc} 
Fit results of SC2H waveforms with a single-pulse function.
SSR and the height parameter $p_2$ obtained in 
the fit are plotted for both of the two photomultipliers (left and right) of SC2H. 
The solid curves are used as boundaries to select single-hit events.
}
\end{figure}

Next, 
TOF between F2 and F4 is analyzed, 
by introducing corrections for time-walk effects and dependence on ion-optical variables. 
The corrected TOF spectrum for  
the data set at  
the scaling factor $f=0.980$ 
is presented in Fig.~4 as an example. 
The unshaded spectrum shows the total recorded events, 
where a peak for the deuteron  
is observed above  
a constant component  
due to the multiple-hit proton background.  
The achieved time resolution for the deuteron peak is $\sigma = 1.7 \times 10^{2}$~ps. 
The shaded spectrum
displays the single-hit events selected by the waveform analysis,
demonstrating efficient rejection of the multiple-hit background by 2--3 orders of magnitude.  
Finally,  
events within the dashed lines ($\pm 5 \sigma$ region around the peak) are selected
as deuteron events  
to further reject remaining proton background.

\begin{figure}[hbtp]
\centering \includegraphics[width=85.0mm]{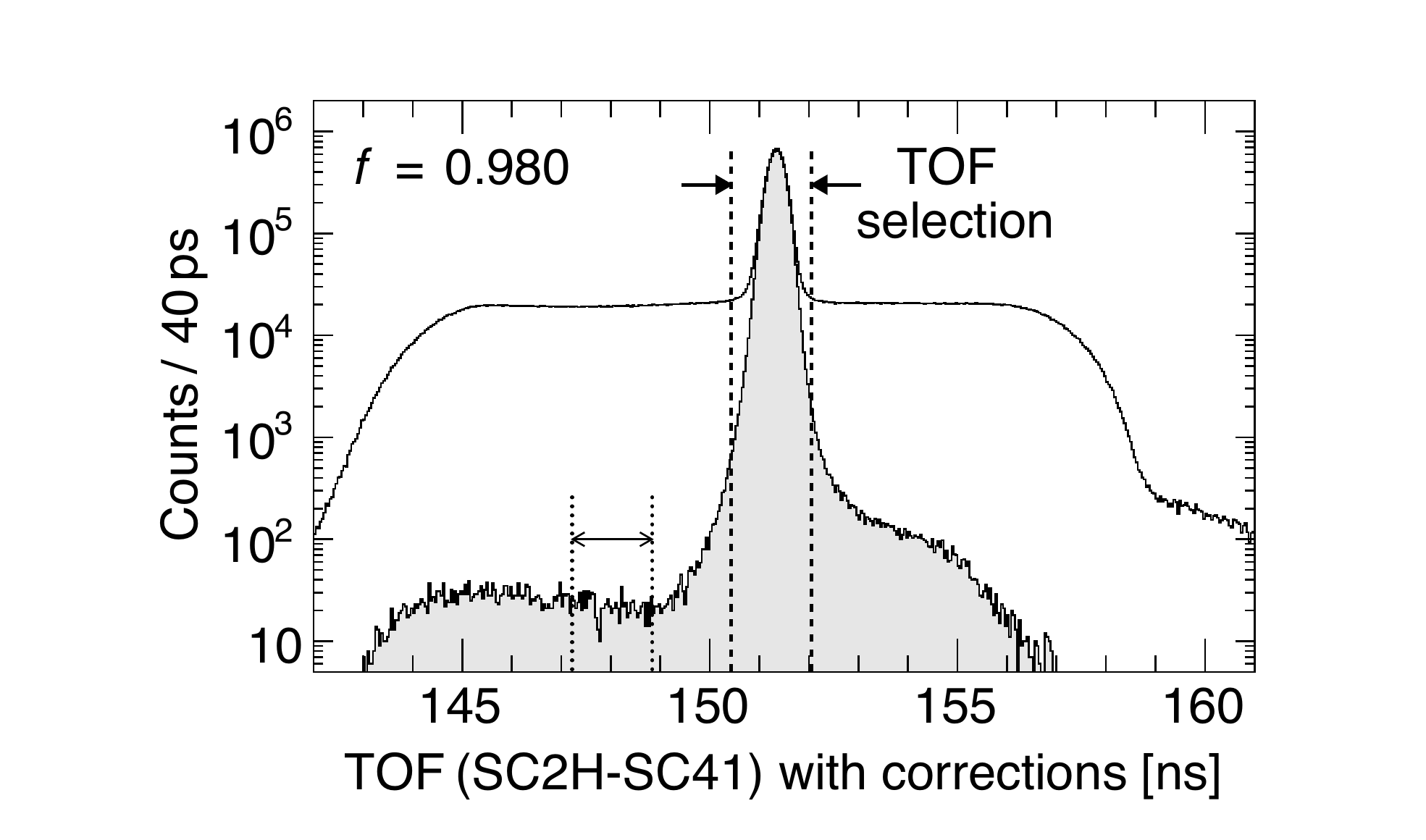}
\caption{\label{fig_pid_tof} 
Corrected TOF spectra between F2 and F4 at the FRS scale factor of $f=0.980$. 
The unshaded histogram shows the total number of events, 
while the shaded one corresponds to the single-hit events selected by the SC2H waveforms. 
The dashed lines indicate the $\pm 5 \sigma$ region around the deuteron peak used for the TOF selection.
The region between the dotted lines is used to estimate the proton contamination fraction.
}
\end{figure}

The efficiency of the deuteron identification 
is discussed in three steps as follows.  
First, properly measured times and
pulse heights of the SC2H and SC41 signals
and reconstructed tracks at F4 
are required for the analysis of the particle identification.
This  
first condition leads to ${0.6\text{--}0.9}$\% rejection of the deuteron. 
Second,    
rejection of the deuteron 
by selecting 
the single-hit waveform is considered using  
a TOF spectrum for those events rejected 
in the waveform selection.
The spectrum shows a small enhancement at the TOF value for the deuteron, 
indicating 2--3\% rejection. 
Third, probability of deuteron rejection
by the TOF selection is 
estimated to be 0.2--0.3\% from the tail structure
of the deuteron peak observed in the TOF spectrum. 
The longer tail on the right side is 
due to 
accidental particles almost coincident with the deuteron at F2
where the timing information measured by a leading-edge discriminator
deviated only to the earlier side.
Combining the above three contributions,
the deuteron identification efficiency is evaluated to be 
96--97\% for all the data sets of the ${}^{12}$C($p$,$d$) reaction. 

The contamination of background protons in 
events with identified deuteron 
is evaluated  
in the TOF spectrum with the single-hit selection.
The spectrum shows an 
almost constant background on the shorter TOF side of the deuteron peak,
where a tail structure of the peak is not significant. 
The amount of
the contamination in the TOF window (dashed lines)
can be estimated by integrating the constant region between the dotted lines with the same interval.
The contamination fraction thus evaluated is $\sim 2 \times 10^{-4}$, 
making only a negligible contribution in the subsequent spectral analysis.

\subsection{Momentum analysis \label{ss_momentum_analysis}}

The deuteron momentum is obtained from the reconstructed track 
by MWDCs at the F4 dispersive focal plane. The momentum $P_d$ can be written as
\begin{equation} P_d = P_{\mathrm{FRS}}  (1+ \delta), \end{equation}
where $\delta$ denotes a momentum deviation relative 
to the FRS central momentum, 
$P_{\mathrm{FRS}} = f \cdot 2828.0$~MeV/$c$. 
The deviation 
$\delta$ can be derived from the horizontal track (position $X$ and angle $X'$) 
regardless of the scaling factor $f$, since 
ion-optical properties remain unchanged
by scaling the central momentum in a small range.

In order to obtain a calibration function converting 
a track ($X$, $X'$) to $\delta$,  
the mono-energetic deuteron from  
the elastic $\mathrm{D}(p,d)p$ reaction 
at the proton energy of 1.6~GeV
is analyzed.
A deuteron emitted in this reaction has a momentum of $2828.0$~MeV/$c$,
corresponding to a deviation of $\delta = 1/f -1$ for the FRS scaling factor $f$.
Thus, the ion-optical response for $\delta$ between $-2\%$ and $2\%$ 
can be evaluated from the calibration settings 
listed in Table~\ref{table_measurement_summary}. 
Examples of  the reconstructed horizontal position $X$ and angle $X'$ are shown in the left panel of Fig.~5, overlaid for $f=0.980, 0.990, 1.000, 1.010, 1.020$.   
An elastic-scattering locus for each scaling factor is observed above 
a continuum from reactions with carbon.  
We fit the position $X$ by a polynomial function of both $X'$ and $\delta$ using all the data sets,
and thereby construct a calibration function for $\delta$, as demonstrated in the right panel.

An uncertainty associated with the calibration of $\delta$ is 
derived
from deviations between the repeated measurements.
Thus the estimated systematic error of $\delta$ 
is $0.02\%$ in a region of $|X'|<18$~mrad, which 
is later used in the analysis of the ${}^{12}$C($p$,$d$) reaction.

\begin{figure}[hbtp]
\centering \includegraphics[width=85.0mm]{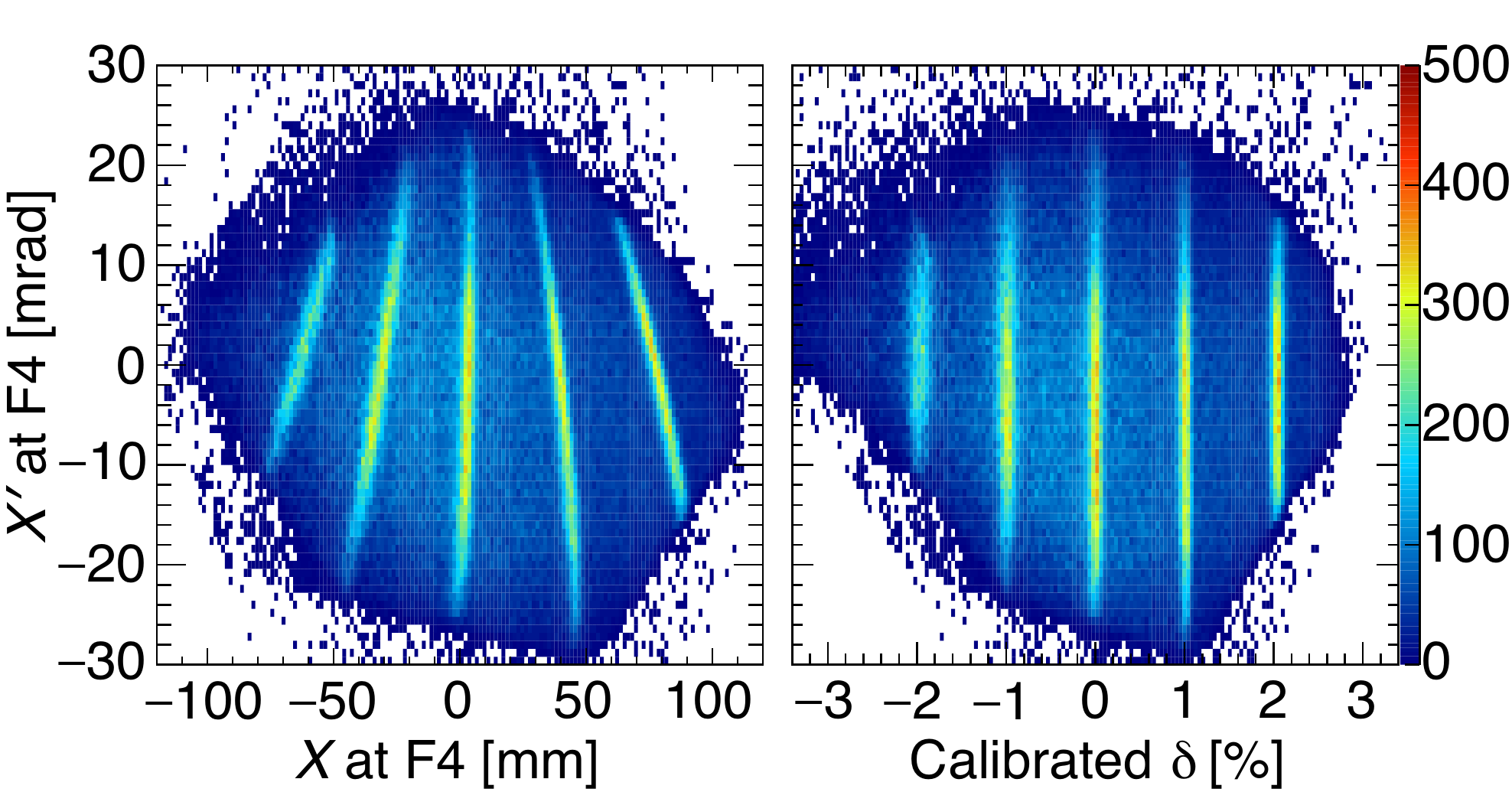}
\caption{Calibration data with the elastic proton-deuteron scattering   
for the FRS scale factors $f=0.980$, 0.990, 1.000, 1.010, 1.020. 
(Left panel) Horizontal position $X$ and angle $X^\prime$ at F4 reconstructed by MWDCs. 
(Right panel) Calibrated $\delta$, the momentum deviation relative to the FRS central momentum, 
is shown against the angle $X^\prime$.
}
\end{figure}

\subsection{Excitation energy \label{ss_excitation_energy}}
\subsubsection{Calculation}
The excitation energy of ${}^{11}$C 
has been calculated
from the proton kinetic energy $T_p$ and the deuteron momentum $P_d$.
First, the relativistic energies  
of the proton ($E_p'$)   
and the deuteron ($E_d'$) 
are calculated
at the center of the reaction target, 
after correcting the energy losses, as 
\begin{eqnarray}
E_p'  &=&   T_p + M_p c^2 - \Delta{E}_p, \\ 
E_d'  &=&  \sqrt{ P_d^2 c^2 + M_d^2 c^4} + \Delta{E}_d. 
\end{eqnarray}
$\Delta{E}_p$ and $\Delta{E}_d$ represent the  
energy losses in half of the target thickness
based on a calculation with {\sc{atima}} \cite{ATIMA}.
The missing mass $M_X$ in the ${}^{12}$C($p$,$d$)$X$ reaction 
is 
then obtained by
\begin{equation}
M_{X} = \sqrt{ \left( M_{^{12}\mathrm{C}} + \frac{E'_p  - E'_d}{c^2} \right)^2 - \left(
\frac{P_p' - P_d'}{c}\right)^2 },
\label{eqs_mm}
\end{equation}
where $M_{^{12}\mathrm{C}}$ is the mass of $^{12}\mathrm{C}$, 
and $P_p'$ and $P_d'$ are the proton and deuteron 
momenta corresponding to $E_p'$ and $E_d'$, respectively. 
Here, the reaction angle is assumed to be  
0$^\circ$ 
because of the ion-optical restrictions on angle reconstruction. 
Finally, 
the excitation energy $E_{\mathrm{ex}}$ is defined
relative to the $\eta^\prime$ production threshold $E_0=957.78$~MeV  as
\begin{equation}
E_{\mathrm{ex}}-E_{0} = \left({M_X} - M_{^{11}\mathrm{C}} - M_{\eta^\prime} \right) c^2,	
\end{equation}
where $M_{^{11}\mathrm{C}}$ and $M_{\eta^\prime}$ denote 
the ${^{11}\mathrm{C}}$ and $\eta^\prime$ masses, respectively.

\subsubsection{Systematic error \label{sss_Eex_systematic_error}} 
The systematic errors in the excitation-energy calculation
have been evaluated
by considering the following three sources.

\textit{(a) Beam energy.} 
Uncertainties in the absolute beam energies 
($2499.1 \pm 2.0$~MeV and $1621.6 \pm 0.8$~MeV) 
cause systematic errors of 1.9~MeV and 0.8~MeV in the excitation energy, respectively.
Note here that the lower beam energy also affects the excitation energy, 
as the deuteron momentum in the ${\mathrm{D}}(p, d)p$ calibration at 1.6~GeV
is used to define the FRS central momentum. 
Considering that the two errors are mostly correlated 
because of their common source  
in the circumference of the SIS-18 synchrotron, 
we obtain the combined systematic error of 1.4 MeV in the excitation energy.

\textit{(b) Reaction angle.} 
A systematic error originating in 
the uncertainty of the reaction angle, which is 
assumed to be $0^\circ$ in the kinematical calculation of the production and calibration reactions,
is evaluated to be 0.8 MeV in 
the excitation energy.
The possible maximum systematic error 
caused by neglecting the finite 
reaction-angle distribution ($\lesssim 1^\circ$) in the acceptance of the FRS
is adopted. 

\textit{(c) Optics calibration.} 
The systematic error 
of the relative momentum deviation $\delta$
has been evaluated to be $0.02\%$. 
This corresponds to an error of 0.5~MeV 
in the scale of the excitation energy. 

By taking a square root of the quadratic sum of all the above contributions,
a total systematic error in the excitation energy is estimated to be 1.7~MeV.

\subsubsection{Experimental resolution \label{sss_Eex_resolution}}

\begin{figure}[hbtp]
\centering \includegraphics[width=85.0mm]{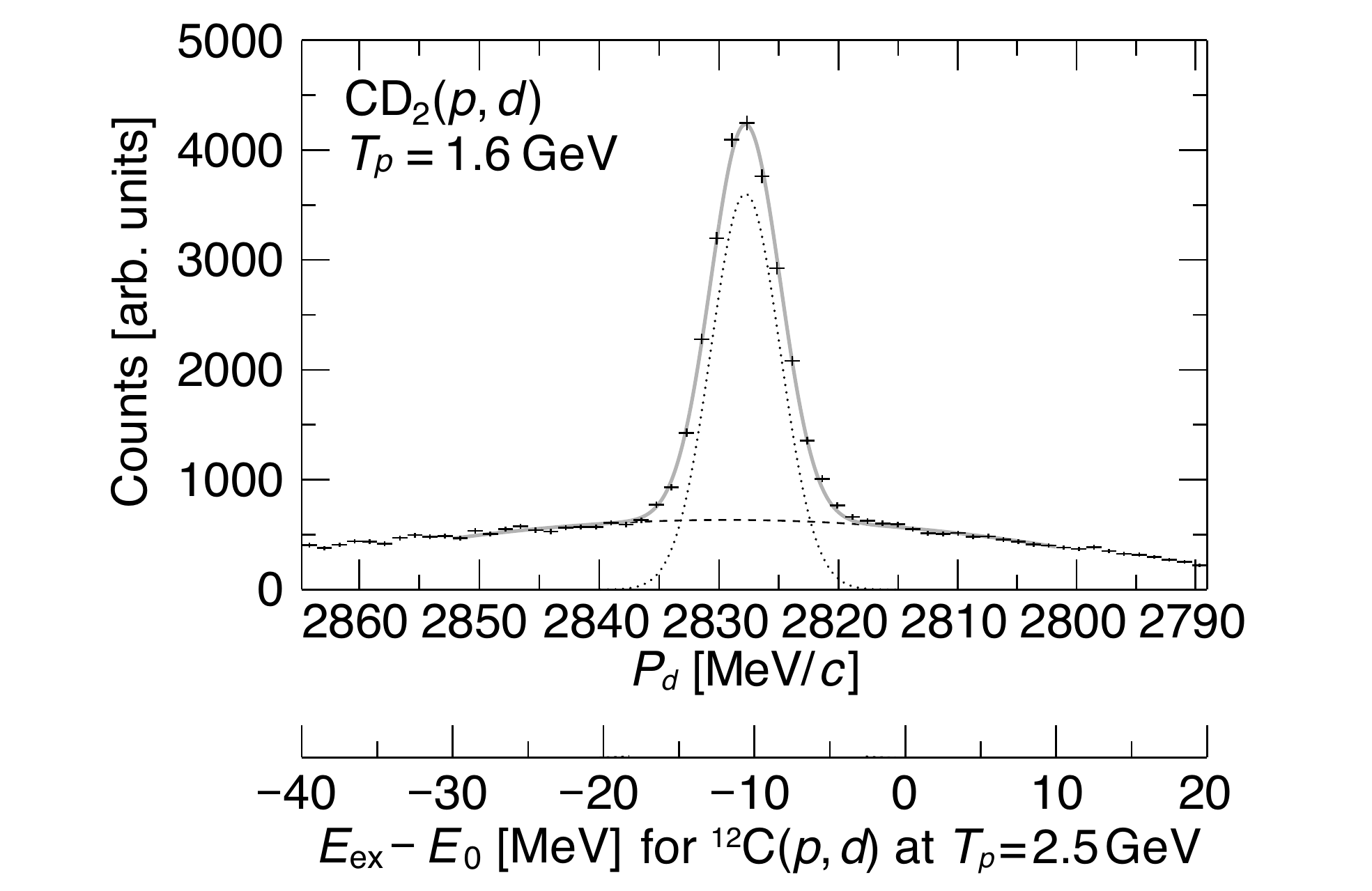}
\caption{\label{fig_resolution} 
Momentum spectrum of deuterons   
in the $\mathrm{CD}_2 (p,d)$ calibration measurement with
the 1.6 GeV proton beam.
The  
lower axis shows the excitation 
energy for the production run 
corresponding to the deuteron momentum. 
The gray solid line displays a fit with a function 
consisting of 
a Gaussian peak (dotted line) 
and a second-order polynomial (dashed line). 
}
\end{figure}

The experimental resolution  
is evaluated using the mono-energetic deuterons  
in the $\mathrm{D}(p,d)p$ calibration
at 1.6 GeV. 
Figure~\ref{fig_resolution} shows a spectrum of the deuteron momentum
at the FRS scaling factor of $f=1.000$
analyzed in the same procedure as the production runs. 
The spectrum is fitted well with a function given by the gray solid line, 
summing a Gaussian function for the elastic peak (dotted line) 
and a second-order polynomial for the carbon contribution (dashed line). 
The Gaussian component yields 
the overall momentum resolution
of $2.79 \pm 0.09$~MeV/$c$ ($\sigma$), where the uncertainty includes 
deviations between the  
different data sets. 
This resolution corresponds to an energy resolution of 
$\sigma_{\mathrm{cal}} = 2.20 \pm 0.07$~MeV 
in the scale of $E_{\mathrm{ex}}$ shown 
in the lower axis.

In order to evaluate the experimental resolution in the production runs
from the above estimate for the calibration runs,
two corrections for the different target and beam energy
in the calibration measurements are necessary. 
First, a Monte-Carlo simulation 
based on {\sc{atima}} \cite{ATIMA} shows that the energy straggling in the targets 
makes contributions of
$\sigma^{\mathrm{targ}}_{\mathrm{prod}} = 1.15$~MeV and 
$\sigma^{\mathrm{targ}}_{\mathrm{cal}} = 0.39$~MeV 
in the production and calibration runs, respectively, 
in the scale of $E_\mathrm{ex}$. 
Second, an estimated relative momentum spread 
of the beams of $\leq 1.7\times 10^{-4}$ 
accounts for 
resolutions of $\sigma^{\mathrm{beam}} _{\mathrm{prod}} \leq 0.52$~MeV in the production runs 
and  
$\sigma^{\mathrm{beam}} _{\mathrm{cal}} \leq 0.34$~MeV in the calibration 
in the scale of $E_\mathrm{ex}$. 
By introducing the corrections quadratically {\textit{i.e.,}} $\sigma_E ^2 = \sigma_{\mathrm{cal}}^2 + \{ (\sigma^{\mathrm{beam}} _{\mathrm{prod}})^2 - (\sigma^{\mathrm{beam}} _{\mathrm{cal}})^2\}+ \{ (\sigma^{\mathrm{targ}} _{\mathrm{prod}})^2 - (\sigma^{\mathrm{targ}} _{\mathrm{cal}})^2\}$, 
the excitation-energy resolution in the production measurements 
is obtained as
$\sigma_E = $ 2.4--2.6~MeV.

\subsection{Normalization of spectra \label{ss_normalization}}

Excitation-energy spectra  
need to be corrected for 
the acceptance of the spectrometer,
which can be expressed as a function of the momentum parameter $\delta$. 
For spectral analysis, events with the horizontal angle at F4 
in $|X'| < 18$~mrad and $\delta$ in $ 0 \% \leq  \delta \leq  1.5\%$
are selected, where the acceptance curve is assumed to be linear.  
The slope of the acceptance function is then deduced by two methods: 
(i) measuring the $\delta$-dependence of deuteron yields at one fixed absolute momentum
 by scaling the FRS magnetic fields,  
and (ii) a Monte Carlo simulation of the ion-optical transport 
in the FRS using the code {\sc{mocadi}} \cite{Iwasa1997}. 
These two estimations result in a consistent acceptance curve of 
$A(\delta) \propto 1+(0.07\pm0.03)\delta/\%$, 
and the excitation spectra at the seven FRS settings are thus corrected with this function.

To normalize the absolute scale of the double differential cross section, 
we analyze a short production run with the SEETRAM detector directly monitoring the beam intensity.
The double differential cross section is calculated by 
\begin{equation}
\left( \frac{d^2 \sigma}{d \Omega dE} \right)_{\text{\!lab}} \!
= \frac{ 
({dY}/{dE})_{\mathrm{ref}}
}{ N_p \cdot n_{t}  \cdot \Delta \Omega_{\text{ref}} \cdot \varepsilon}\label{eq_diff_cross_section}
\end{equation}
at one reference energy $E_{\mathrm{ex}}-E_0 = -7.0~\mathrm{MeV}$,
where the solid angle 
covered by the FRS is separately evaluated as 
$\Delta \Omega_{\mathrm{ref}} = 1.16 \pm 0.13 $~msr. 
$({dY}/{dE})_{\mathrm{ref}}$ is the measured yield density per unit excitation energy at the reference, 
$N_p$ is the number of the incident protons obtained by SEETRAM,
and $n_t$ is the number density of the target.
$\varepsilon$ denotes the overall efficiency estimated to be $(25.2 \pm 0.1)$\%,
taking into account the trigger efficiency ($26.2$\%), 
the deuteron identification efficiency ($96$--$97$\%),
and the tracking efficiency ($99.8$--$99.9$\%). 
As a result, 
the double differential cross section of $5.4 \pm 0.7$~$\mu$b/(sr MeV) 
is obtained at the reference energy.

Spectral normalization is then performed as follows. 
First, 
the acceptance-corrected excitation spectrum
at the FRS scale factor of $f=0.990$ is normalized, 
according to the analyzed cross section 
at the reference energy $E_{\mathrm{ex}}-E_0 = -7.0~\mathrm{MeV}$.
Next, the spectra at the neighboring FRS settings ($f=0.985$ and $1.000$) 
are scaled so that they have 
consistent 
overlap in the common energy region with the $f=0.990$ spectrum. 
Spectra at the other settings are sequentially normalized in the same manner. 

The differential cross section 
of the elastic $\mathrm{D}(p,d)p$ reaction 
at  
2.5~GeV  
is analyzed as well in a similar way
in order to confirm the analyses on the beam intensity at this energy
and on the normalization of the cross section.
The obtained value is 
$ 0.98 \pm 0.11$~$\mu$b/sr in the center-of-mass frame,
which is consistent with known cross sections around this energy,  
$1.10 \pm 0.10 (\mathrm{stat}) \pm 0.09 (\mathrm{syst})$~$\mu$b/sr at 2.4~GeV
and $1.18 \pm 0.03 (\mathrm{stat}) \pm 0.09 (\mathrm{syst})$~$\mu$b/sr at 2.55~GeV
\cite{Berthet1982}, 
within the experimental errors.


\section{Results \label{s_results}}

\begin{figure}[hbtp]
\centering \includegraphics[width=85.0mm]{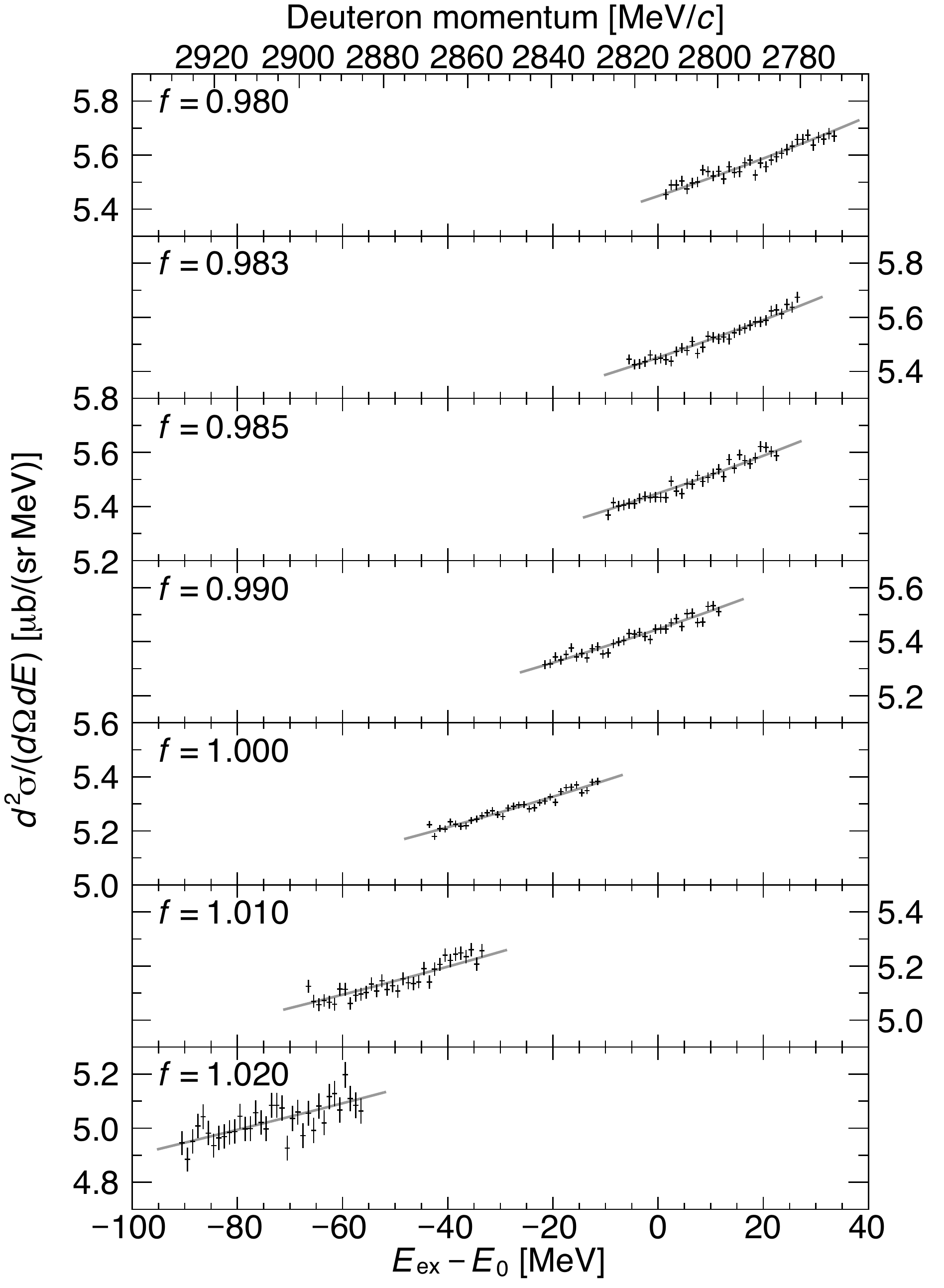}
\caption{\label{fig_spectra_all} 
Excitation-energy spectra of $^{11}$C measured in the $^{12}$C($p$,$d$) reaction
with seven momentum settings ($f=0.980$--$1.020$) of the FRS. 
The lower abscissa is the excitation energy $E_{\mathrm{ex}}$ relative to 
the $\eta^\prime$ production threshold $E_0 = 957.78$~MeV, 
and the upper axis shows the corresponding deuteron momentum. 
The gray solid curves display a third-order polynomial simultaneously fitted to the seven spectra.
}
\end{figure}

Excitation-energy spectra of $^{11}$C obtained with the $^{12}$C($p$,$d$) reaction 
are presented in Fig.~\ref{fig_spectra_all} 
for the seven momentum settings of the spectrometer. 
The excitation energy $E_{\mathrm{ex}}$ 
relative to 
the $\eta^\prime$ production threshold $E_0=957.78$~MeV 
is shown in the lower horizontal axis, while the scale 
of the corresponding deuteron momentum is given in the upper one.
The systematic error associated with the excitation energy 
has been estimated to be 1.7~MeV.
The  
ordinate  
gives
the double differential cross section,
with an  
uncertainty of $\pm 13\%$ on the absolute scale.

\begin{figure}[hbtp]
\centering \includegraphics[width=85.0mm]{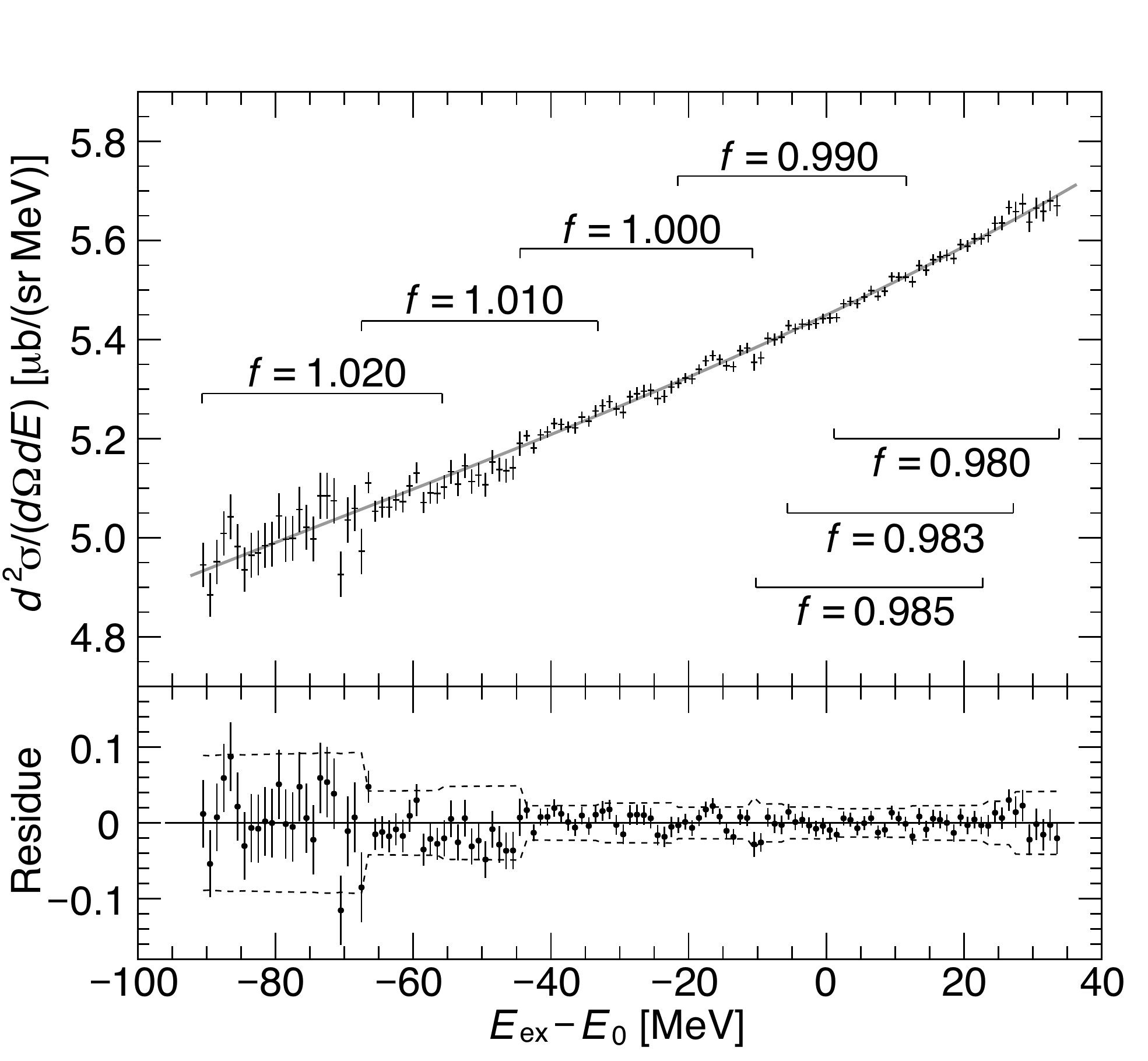}\caption{\label{fig_combined_spectrum} 
(Top panel) Combined spectrum of the $^{11}$C excitation energy 
$E_{\mathrm{ex}}$ near the $\eta^\prime$ production threshold $E_0 = 957.78$~MeV.  
A third-order polynomial fit is given by the solid gray curve. 
The horizontal bars indicate
excitation-energy regions covered by each FRS momentum setting.
(Bottom panel) Residues of the polynomial fit.
The dashed lines display envelopes of 2 standard deviations.
}
\end{figure}

These seven spectra are combined into one spectrum
by averaging the data points of  
different FRS settings at each excitation energy.
The resulting    
excitation spectrum 
is shown in Fig.~\ref{fig_combined_spectrum} (top panel).
Note that this averaging  
reduces the degrees of freedom originating in
the relative spectral normalization between the neighboring settings.
Therefore, the following analyses are performed 
for both the individual spectra (Fig.~\ref{fig_spectra_all})
and the combined one (Fig.~\ref{fig_combined_spectrum}), where 
only minor differences  
are found  
as shown
later.

The obtained spectra show no distinct narrow structure,
although a good statistical sensitivity
at a level of $<1\%$ is achieved 
together with a sufficiently good 
experimental resolution of
$\sigma_E =$ 2.4--2.6~MeV.
The spectra exhibit a smooth increase from about 4.9 to 5.7~$\mu$b/(sr MeV) 
in the measured region of the excitation energy. 
This continuous component can be understood 
within an order of magnitude
by quasi-free meson production processes 
$pN \rightarrow dX$ ($X=2\pi, 3\pi, 4\pi, \omega$), 
where $N$ denotes a nucleon in the target nucleus, 
as simulated in Ref.~\cite{Itahashi2012}. 

The spectra are fitted by a third-order polynomial function 
over the whole measured energy region. 
The seven spectra in Fig.~\ref{fig_spectra_all} are simultaneously fitted, 
by sharing the polynomial parameters between the settings.
In addition, a multiplying factor is introduced as a free parameter 
to each spectrum except for the one at $f=0.990$
in order to
take into account a possible error correlation
with the relative normalization of the spectra.
The fit results are shown by the gray solid curves in both Figs.~\ref{fig_spectra_all} 
and~\ref{fig_combined_spectrum}.  
$\chi^2/(\mathrm{n.d.f.})$ 
is  
221/225 in Fig.~\ref{fig_spectra_all}
and 
125/121 in Fig.~\ref{fig_combined_spectrum}, 
where n.d.f.~denotes the number of degrees of freedom. 
Residues of the fit are also displayed in Fig.~\ref{fig_combined_spectrum} (bottom panel)
with envelopes of 2 standard deviations.

We determine upper limits for the formation cross section of $\eta^\prime$-mesic nuclei.
Here, a Lorentzian function  
at an excitation energy $E_{\mathrm{ex}}$ with a width $\Gamma$ (FWHM)
is tested as a signal shape.
The measured spectrum is assumed to be described
by the following function:
\begin{eqnarray}
f(E;E_{\mathrm{ex}},\Gamma)= \left(\frac{d\sigma}{d\Omega}\right) \cdot \mathrm{Voigt}(E; E_{\mathrm{ex}},\Gamma,\sigma_E)  \notag \\
+ (p_0+p_1 E + p_2 E^2 +p_3 E^3)	.   
\end{eqnarray}
The first term includes 
 a Voigt function, which is the Lorentzian function 
folded by a Gaussian function accounting for the experimental resolution ($\sigma_E$). 
The signal cross section to be tested 
is represented by ($\frac{d\sigma}{d\Omega}$). 
The second term is the third-order polynomial accounting for the continuous component. 
The combined spectrum (Fig.~\ref{fig_combined_spectrum}) is fitted 
by this function within a region of $\pm 35$~MeV around the Lorentzian center,  
by treating both the signal cross section and 
the polynomial coefficients as free parameters. 
The upper limit of the cross section 
at the 95\% confidence level is then determined
by assuming a Gaussian probability density function  
based on the fit result and normalizing it 
in the physical non-negative region ($\frac{d\sigma}{d\Omega} \geq 0$).

\begin{figure}[hbtp]
\centering \includegraphics[width=85.0mm]{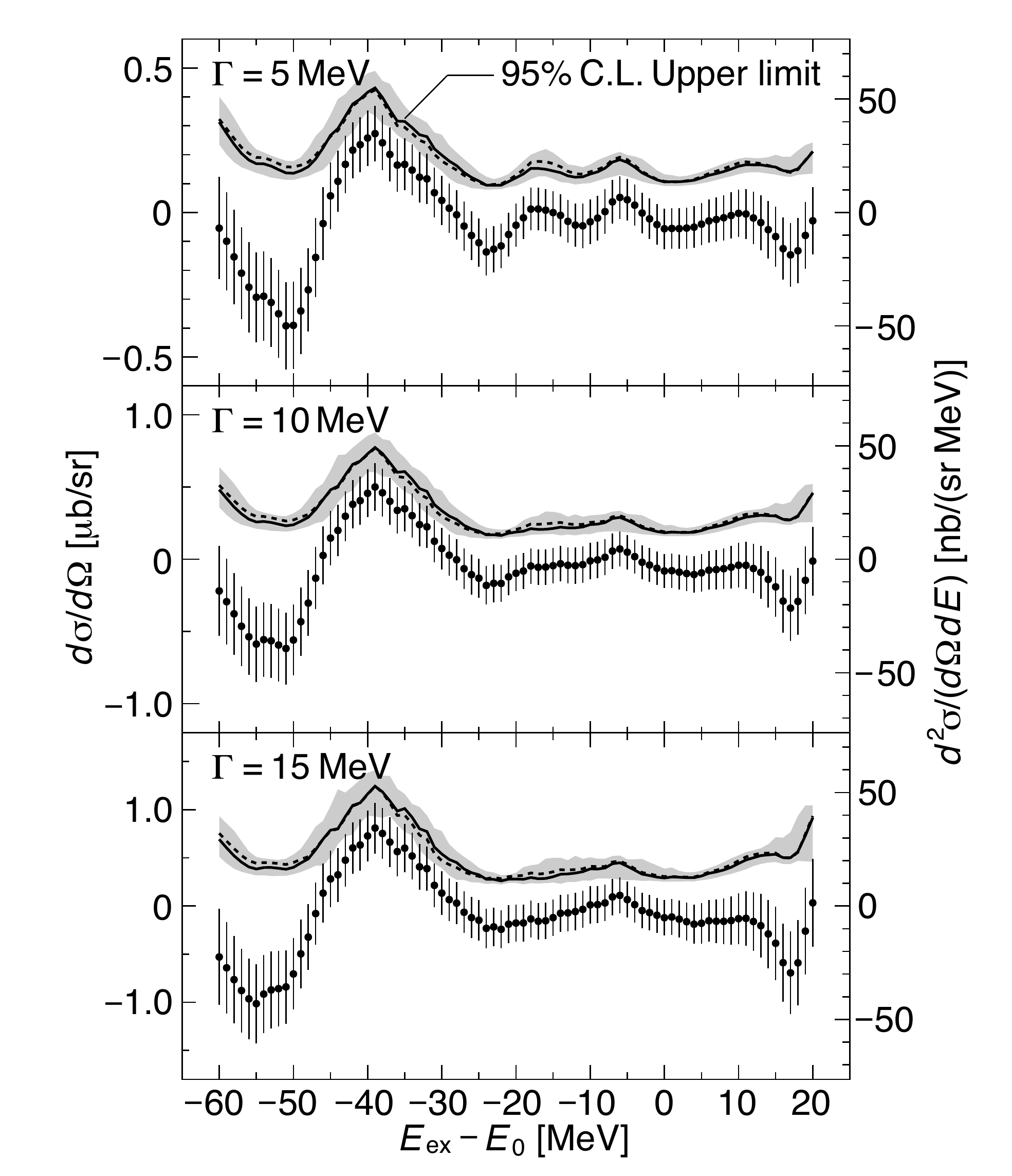} 
\caption{\label{fig_upper_limit} 
Fitted values (solid dots)  
and resulting 95\% C.L. upper limits (solid curves) 
of the Lorentzian-shaped formation cross section of 
$\eta^\prime$-mesic nuclei 
plotted as a function of the assumed peak position $E_{\mathrm{ex}}$ 
for the widths of $\Gamma = 5, 10, 15$~MeV.
Upper limits evaluated by simultaneous 
fitting of the seven spectra (Fig.~\ref{fig_spectra_all}) 
are shown by the dashed curves.
The shaded areas indicate the systematic errors on the upper limits.  
}
\end{figure}

In order to obtain the upper limit 
as a function of the Lorentzian position $E_{\mathrm{ex}}$ and width $\Gamma$, 
the above analysis has been repeated for each set of ($E_{\mathrm{ex}}-E_0$, $\Gamma$) 
in $\{ -60, -59, \dots , +20 \} \times \{ 5, 10, 15 \}$~MeV.  
In Fig.~\ref{fig_upper_limit}, the fitted values and errors
of the Lorentzian cross section are shown by  
the solid dots,
and the resulting upper 
limits are summarized by the solid curves.
These values are given 
in the differential cross section $d\sigma/d\Omega$ by the left ordinate
and in the Lorentzian peak height $d^2\sigma/(d\Omega dE)$ by the right one. 
Moreover, analysis based on simultaneous fitting of the seven spectra (Fig.~\ref{fig_spectra_all}) 
is conducted in order to check   
effects of
possible error correlations with the relative normalization 
of the spectra.
Thus evaluated limits are given by the dashed curves, 
exhibiting no significant difference between the two analysis methods.

To
evaluate systematic errors on the upper limits, 
the following contributions are considered: 
(1) the systematic error on the absolute scale of the measured cross section ($\pm 13\%$), 
(2) the systematic error in the excitation-energy calculation ($\pm 1.7$~MeV),
(3) the uncertainty in the slope of the momentum acceptance,
(4) the uncertainty in the experimental resolution, 
(5) different fit regions (10~MeV wider or narrower), and
(6) a choice of fitting of 
the combined spectrum (Fig.~\ref{fig_combined_spectrum}) 
or simultaneous fitting of the seven spectra (Fig.~\ref{fig_spectra_all}). 
The upper limit is analyzed by changing each of these conditions, and
then the total systematic error on the limit is evaluated by taking 
the square root of the sum of the squared deviations.
In Fig.~\ref{fig_upper_limit}, the thus-evaluated systematic errors 
are displayed by the shaded areas.


\section{Discussion \label{s_discussion}}

\begin{figure}[htbp]
\centering \includegraphics[width=85mm]{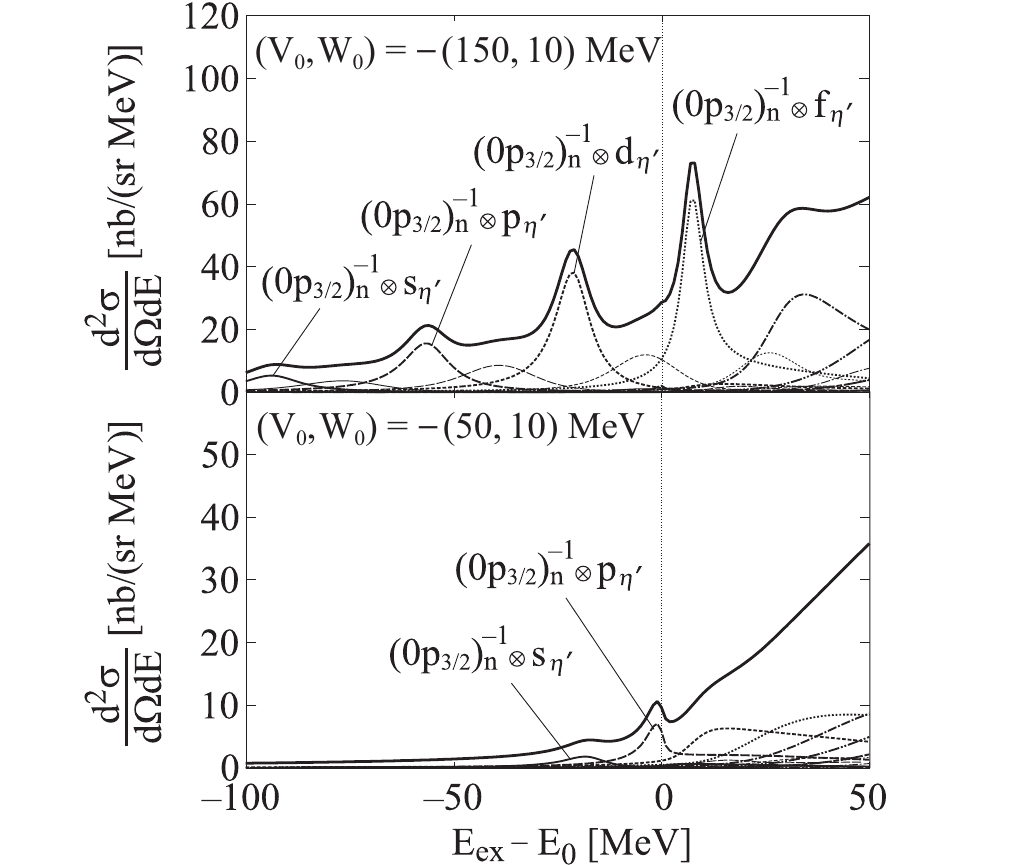}
\caption{\label{fig_theory_spectra}
Theoretically-calculated spectra of 
the $^{12}$C($p, d$)$^{11}$C$\otimes \eta^\prime$ 
reaction at 2.5 GeV \cite{Nagahiro2013}.
The $\eta^\prime$-nucleus potential parameters are taken as 
$(V_0, W_0)=(-150,-10)$~MeV (top)
and $(-50,-10)$~MeV (bottom).
Total formation cross sections are shown by the thick lines,
and major configurations of the $\eta^\prime$-mesic states $\ell_{\eta^\prime}$ 
coupling with the neutron hole states $(n\ell_j)_{n}^{-1}$ are shown by the thin lines.
}
\end{figure}

The 95\% C.L.~upper limits 
for the formation cross section of 
$\eta^\prime$-mesic nuclei have been obtained as a function of the position and width 
of the assumed Lorentzian peak.
The upper limits are particularly stringent near the $\eta^\prime$ production threshold: 
0.1--0.2~$\mu$b/sr for $\Gamma = 5$~MeV,
0.2--0.4~$\mu$b/sr for $\Gamma = 10$~MeV, and
0.3--0.6~$\mu$b/sr for $\Gamma = 15$~MeV.
These are as small as $\sim 20$~nb/(sr~MeV) in the Lorentzian peak height,
and therefore exclude the existence of prominent peak structures 
theoretically expected near the threshold for strongly attractive potentials,
like a peak with $\sim 40$~nb/(sr~MeV) for $V_0 = -150$~MeV
shown in Fig.~\ref{fig_theory_spectra}~(top) \cite{Nagahiro2013}.
On the other hand, the obtained limits
are not in
conflict with
small peak structures predicted for shallow potentials, 
as in Fig.~\ref{fig_theory_spectra}~(bottom) 
for $V_0 = -50$~MeV
where a peak height is $< 10$~nb/(sr~MeV).

In order to make further quantitative comparisons with the theoretical predictions,
we evaluate constraints on the scales 
for the theoretically-calculated formation spectra. 
Here, the 
following function is assumed to describe the measured excitation spectrum,
\begin{eqnarray}
F(E;V_0,W_0) 
&=& \mu \cdot S(E;V_0,W_0,\sigma_E) \notag \\ 
& & + (p_0 + p_1 E + p_2 E^2 +p_3 E^3),
\end{eqnarray}
where $S(E;V_0,W_0,\sigma_E)$ denotes 
the theoretical formation spectrum (e.g., Fig.~\ref{fig_theory_spectra}) \cite{Nagahiro2013,Nagahiro2017}
for the real and imaginary potentials of ($V_0$,$W_0$)
folded by the Gaussian function for the experimental resolution ($\sigma_E$).
The parameter $\mu$ is introduced to test an allowed scale for $S(E;V_0,W_0,\sigma_E)$.
The remaining term is a third-order polynomial for the continuous part of the spectrum. 
A 95\% C.L. upper limit of $\mu$ 
for given ($V_0$,$W_0$)
is analyzed in the similar procedure,
by fitting the measured spectrum with this function 
within the region of $-40$~MeV $\leq E_{\mathrm{ex}} - E_0 \leq +30$~MeV
and assuming a Gaussian probability density function of $\mu$ in the physical region ($\mu \geq 0$).

\begin{figure}[!hbt]
\centering \includegraphics[width=85.0mm]{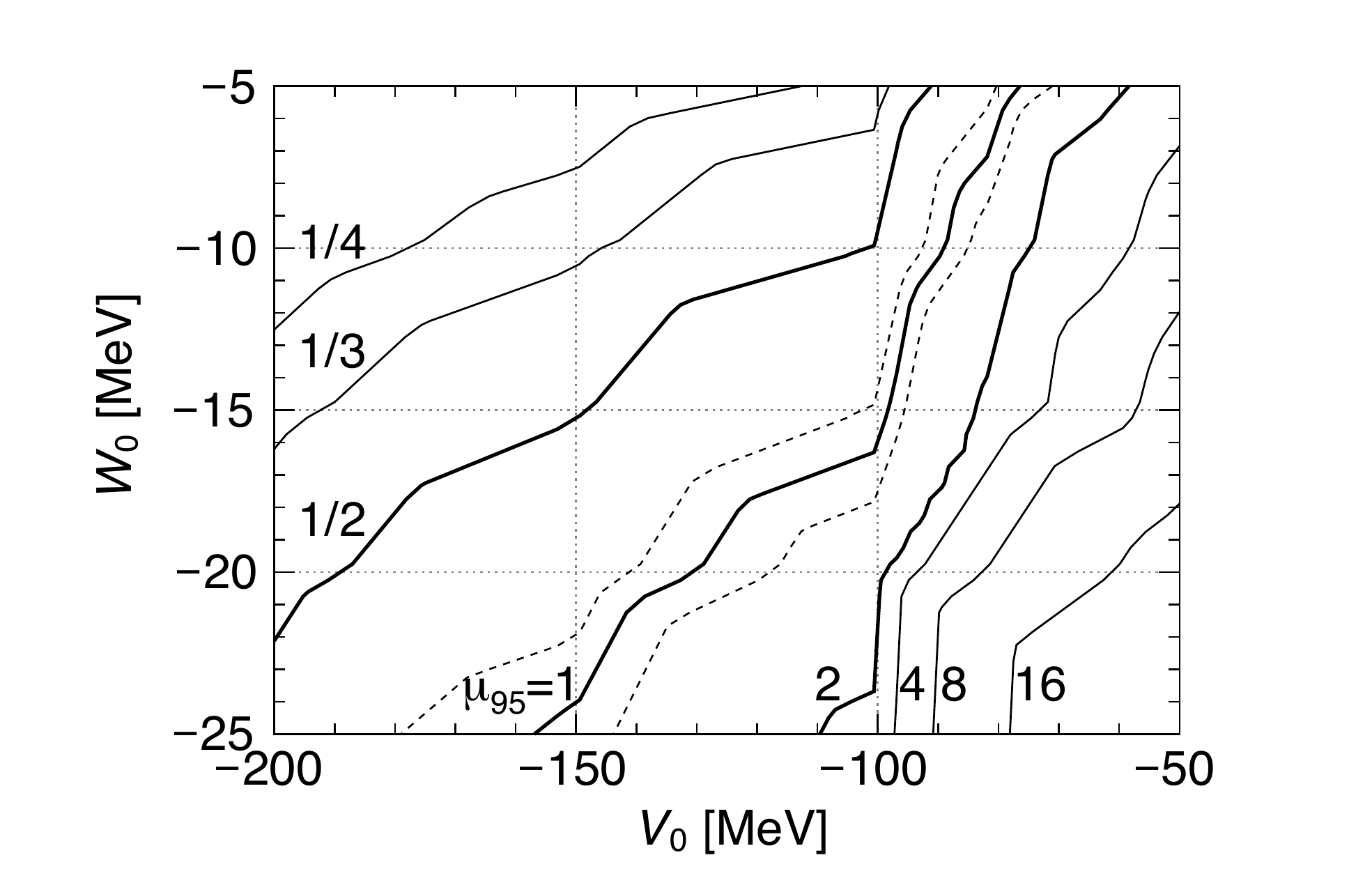}
\caption{\label{fig_VW_result}
A contour plot of $\mu_{95}$ (solid curves), 
the 95\% C.L.~upper limit of the scale parameter $\mu$ 
for the theoretical formation spectra, 
on the real and imaginary potential plane $(V_0,W_0)$.
The limits have been analyzed at the potential sets listed 
in Table~\ref{table_mu_analysis}
and linearly interpolated in-between. 
The systematic errors on the $\mu_{95}=1$ contour are shown by the dashed curves.
The region with $\mu_{95} \leq 1$ is excluded by the present analysis. 
See the text for further explanation.}
\end{figure}

The analysis has been repeated for the potential parameter sets listed in Table~\ref{table_mu_analysis},
including $W_0=-25$~MeV where theoretical spectra were newly calculated \cite{Nagahiro2017}.
Fitted values and resultant 95\% C.L. upper limits  
of the scale parameter $\mu$ are given by $\mu_{\mathrm{fit}}$ and $\mu_{95}$, respectively.
For each
  ($V_0$,$W_0$), the existence of the theoretically-calculated 
  peak structure  with the strength multiplied by $\mu_{95}$ is excluded at 
  the 95\% C.L.
The upper limits $\mu_{95}$ are then linearly interpolated between the calculated potentials,
and presented as a contour plot on the real and imaginary potential plane ($V_0$,$W_0$) in Fig.~\ref{fig_VW_result}.  
Smaller $\mu_{95}$ is deduced for larger $|V_0|$ and smaller $|W_0|$.
Systematic errors on $\mu_{95}$ are estimated by taking into account the same six sources 
as for the Lorentzian upper limits. 
A band of the systematic error on the $\mu_{95}=1$ contour
is shown by the dashed curves.

\begin{table}[tb]
\caption{\label{table_mu_analysis}
Fitted values ($\mu_{\mathrm{fit}}$) and resulting 95\% C.L.~upper limits ($\mu_{95}$)
for the scale of the theoretical formation spectra.  
The analysis is performed at the listed sets of the real and imaginary potentials ($V_0$, $W_0$).}
\begin{ruledtabular}
\begin{tabular}{rrrr}
 \multicolumn{1}{c}{$V_0$} & \multicolumn{1}{c}{$W_0$} & 
 \multicolumn{1}{c}{$\mu_{\mathrm{fit}}$} & \multicolumn{1}{c}{$\mu_{95}$} \\
\multicolumn{1}{c}{(MeV)} & \multicolumn{1}{c}{(MeV)} &  & \\
\hline
$ -50 $ & $ -5 $ & $ 0.04 \pm \phantom{0}1.44 $ & $ 2.85 $ \\ 
$ -50 $ & $ -10 $ & $ 0.22 \pm \phantom{0}2.88 $ & $ 5.78 $ \\
$ -50 $ & $ -15 $ & $ 1.07 \pm \phantom{0}5.29 $ & $ 11.10 $ \\ 
$ -50 $ & $ -20 $ & $ 3.10 \pm \phantom{0}9.11 $ & $ 20.01 $ \\ 
$ -50 $ & $ -25 $ & $ 6.74 \pm 14.75 $ & $ 33.69 $ \\
$ -60 $ & $ -5 $ & $ 0.36 \pm \phantom{0}0.79 $ & $ 1.80 $ \\ 
$ -60 $ & $ -10 $ & $ 0.75 \pm \phantom{0}1.54 $ & $ 3.55 $ \\ 
$ -60 $ & $ -15 $ & $ 1.49 \pm \phantom{0}2.83 $ & $ 6.61 $ \\ 
$ -80 $ & $ -5 $ & $ 0.13 \pm \phantom{0}0.36 $ & $ 0.79 $ \\ 
$ -80 $ & $ -10 $ & $ 0.20 \pm \phantom{0}0.63 $ & $ 1.38 $ \\ 
$ -80 $ & $ -15 $ & $ 0.19 \pm \phantom{0}1.09 $ & $ 2.26 $ \\
$ -100 $ & $ -5 $ & $ -0.24 \pm \phantom{0}0.20 $ & $ 0.27 $ \\ 
$ -100 $ & $ -10 $ & $ -0.32 \pm \phantom{0}0.35 $ & $ 0.50 $ \\ 
$ -100 $ & $ -15 $ & $ -0.43 \pm \phantom{0}0.56 $ & $ 0.86 $ \\ 
$ -100 $ & $ -20 $ & $ -0.60 \pm \phantom{0}0.90 $ & $ 1.41 $ \\ 
$ -100 $ & $ -25 $ & $ -0.85 \pm \phantom{0}1.39 $ & $ 2.22 $ \\
$ -150 $ & $ -5 $ & $ -0.01 \pm \phantom{0}0.10 $ & $ 0.18 $ \\ 
$ -150 $ & $ -10 $ & $ 0.01 \pm \phantom{0}0.15 $ & $ 0.31 $ \\ 
$ -150 $ & $ -15 $ & $ 0.03 \pm \phantom{0}0.23 $ & $ 0.48 $ \\ 
$ -150 $ & $ -20 $ & $ 0.06 \pm \phantom{0}0.35 $ & $ 0.72 $ \\
$ -150 $ & $ -25 $ & $ 0.09 \pm \phantom{0}0.51 $ & $ 1.06 $ \\ 
$ -200 $ & $ -5 $ & $ -0.05 \pm \phantom{0}0.07 $ & $ 0.11 $ \\ 
$ -200 $ & $ -10 $ & $ -0.04 \pm \phantom{0}0.11 $ & $ 0.20 $ \\ 
$ -200 $ & $ -15 $ & $ -0.03 \pm \phantom{0}0.16 $ & $ 0.30 $ \\ 
$ -200 $ & $ -20 $ & $ -0.03 \pm \phantom{0}0.23 $ & $ 0.43 $ \\ 
$ -200 $ & $ -25 $ & $ -0.04 \pm \phantom{0}0.31 $ & $ 0.59 $ 
\end{tabular}
\end{ruledtabular}
\end{table}

In Fig.~\ref{fig_VW_result}, one can exclude a region of the 
potential-parameter set giving $\mu_{95} \leq 1$  
at the 95\% C.L.~within the present comparison with the theoretical calculations.
Note here that the magnitude of the theoretically-calculated spectra 
has an estimated uncertainty of 
a factor $\sim 2$ \cite{Grishina2000, Tanaka2016}, originating 
in the assumed cross section of 30~$\mu$b/sr for the elementary 
reaction $p n \rightarrow d \eta^\prime$ \cite{Itahashi2012}.
Thus, if the theoretical cross sections were overestimated 
by a factor of 2, for example, 
one can reject a potential region with $\mu_{95} \leq 1/2$  
at the 95\% C.L. 
 
\begin{figure}[htb]
\centering \includegraphics[width=85.0mm]{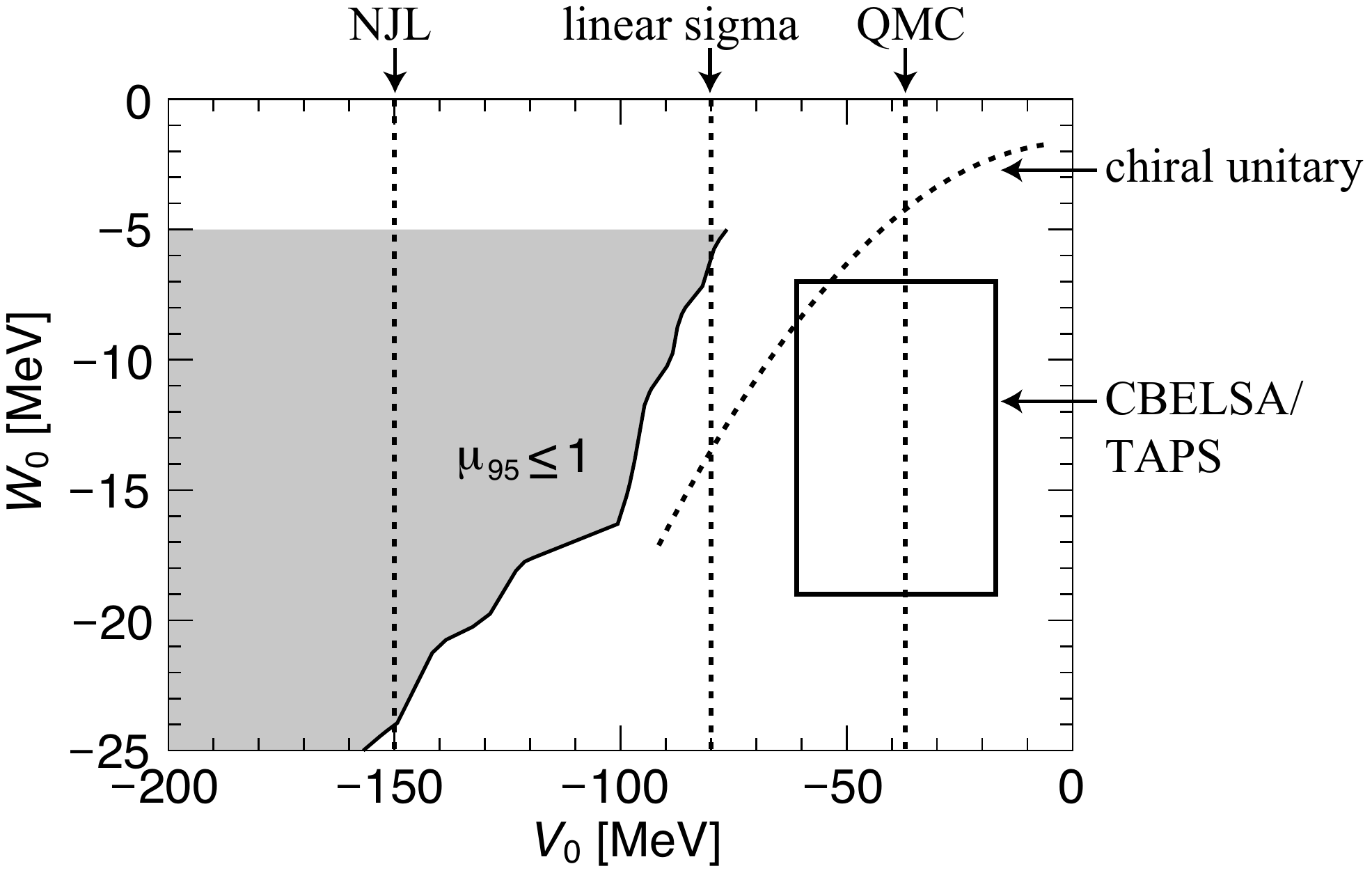}
\caption{\label{fig_VW_overview}
Obtained constraint and currently known information on the $\eta^\prime$-nucleus potential $(V_0 + i W_0)$ at normal nuclear density.  
The shaded region ($\mu_{95} \leq 1$) 
represents the region excluded
within the present analysis.
The rectangular box shows real and imaginary potentials 
evaluated in $\eta^\prime$ photoproduction experiments 
by the CBELSA/TAPS collaboration \cite{Nanova2013, Nanova2016, Nanova2012, Friedrich2016}.
Theoretical predictions based on the NJL model \cite{Costa2005,Nagahiro2006}, 
 the linear sigma model \cite{Sakai2013}, the QMC model \cite{Bass2006}, and the chiral unitary approach \cite{Nagahiro2012}
are indicated by the dashed lines.}
\end{figure}

 
Figure~\ref{fig_VW_overview} summarizes the obtained constraint and 
currently known information on the $\eta^\prime$-nucleus potential.
The shaded region shows the excluded region ($\mu_{95} \leq 1$) in the present analysis.
The rectangular box displays an evaluated region 
by the $\eta^\prime$ photoproduction experiments: 
the real part of $-(39 \pm 7(\mathrm{stat}) \pm 15(\mathrm{syst}))$~MeV  
from the excitation function
and the $\eta^\prime$ momentum distribution \cite{Nanova2013, Nanova2016}, and the imaginary part 
of $-(13 \pm 3(\mathrm{stat}) \pm 3(\mathrm{syst}))$~MeV
from 
the transparency ratios \cite{Nanova2012, Friedrich2016}.
Theoretical expectations with 
the NJL model ($V_0=-150$~MeV) \cite{Costa2005,Nagahiro2006}, the linear sigma model ($V_0=-80$~MeV) \cite{Sakai2013}, 
the QMC model ($V_0=-37$~MeV) \cite{Bass2006}, and the chiral unitary approach \cite{Nagahiro2012}
are shown by the dashed lines.

Here, 
a strongly attractive potential of the order of 
$V_0 \approx -150$~MeV, as predicted by the NJL model, 
is rejected within the present analysis for the  
region of the imaginary potential of $|W_0| \leq 24$~MeV. 
The current experiment has very limited sensitivity 
in a shallower potential region 
where some small peak structures are predicted in the theoretical formation spectra \cite{Nagahiro2013}, 
as shown in Fig.~\ref{fig_theory_spectra} (bottom), for example.
Therefore,
an improvement
of the experimental sensitivity 
is necessary for further investigating the existence of 
$\eta^\prime$-mesic nuclei.

\begin{figure}[htb] 
\centering \includegraphics[width=85.0mm]{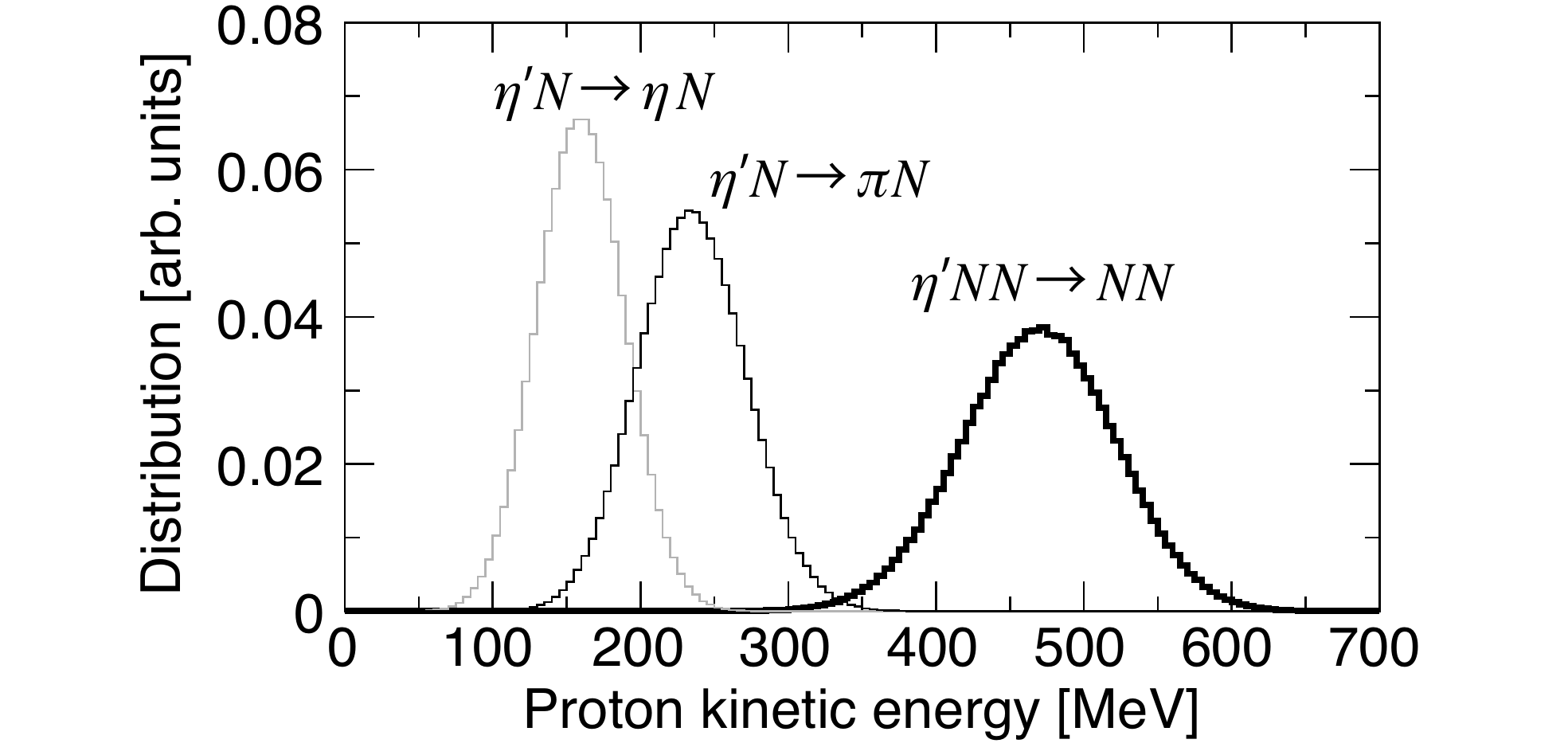}
\caption{\label{fig_momentum_simulation}
Simulated kinetic energy distributions of protons emitted in
the decay of $\eta^\prime$-mesic nuclei.
Three decay modes are considered: $\eta^\prime  N \rightarrow \eta N$ (thin gray), 
$\eta^\prime  N \rightarrow \pi N$ (thin black), and
$\eta^\prime N N \rightarrow N N$  (thick).
The integral of each distribution is normalized to unity.}
\end{figure}

One of the possible approaches for the next step is 
a semi-exclusive measurement by
simultaneously detecting the forward deuteron in the $^{12}$C($p,d$) reaction 
for missing-mass spectrometry
and decay particles from $\eta^\prime$-mesic nuclei for event selection. 
A large amount of the continuous background dominating 
the present spectrum in Fig.~\ref{fig_combined_spectrum},
which is understood as quasi-free multi-pion production,
will be suppressed by tagging the decay particles.
As discussed in Ref.~\cite{Nagahiro_NPA2013}, major decay modes of 
the $\eta^\prime$-mesic nuclei 
are expected to be one- and two-nucleon absorption:
$\eta^\prime  N \rightarrow \eta N$, 
$\eta^\prime  N \rightarrow \pi N$, and 
$\eta^\prime N N \rightarrow N N$.
Among them the two-nucleon absorption process 
has a distinguishing feature in the emitted 
proton (or neutron) energy of $\sim$ 300--600~MeV, 
as simulated in Fig.~\ref{fig_momentum_simulation} assuming 
the Fermi motion of nucleons in the nucleus \cite{Moniz1971}. 
A simulation based on an intra-nuclear cascade model \cite{Nara2000} 
has shown that 
the signal-to-background ratio will be increased by two orders of magnitude
compared to the present experiment
by selecting energetic protons in the backward angular range ($\theta^{\mathrm{lab}}_p \geq 90^\circ$) in the laboratory \cite{HigashiThesis}.

\begin{figure}[htb]
\centering \includegraphics[width=85.0mm]{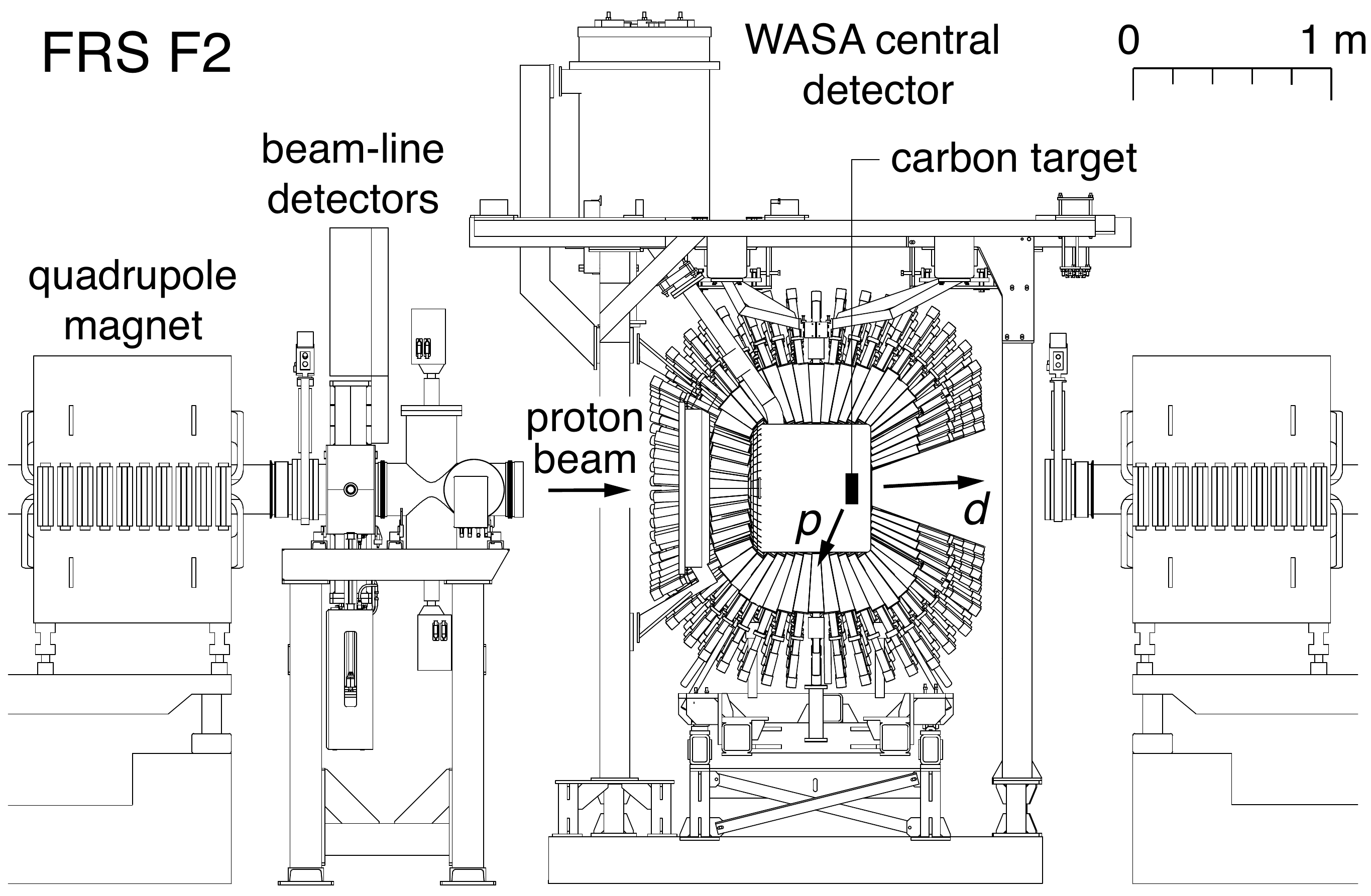}
\caption{\label{fig_setup_semiexclusive}
A schematic view of
an experimental setup for 
the semi-exclusive measurement. 
A 2.5~GeV proton beam impinges on a carbon target. 
The forward deuteron is momentum-analyzed by the downstream FRS section. 
The proton emitted backward in the decay of 
$\eta^\prime$-mesic nuclei is identified by the WASA central detector \cite{WASA_CELSIUS, WASA_COSY}.}
\end{figure}

The semi-exclusive measurement will be performed in the near future.
This experiment is feasible with the FRS at GSI and the next-generation Super-FRS \cite{Geissel2003} at FAIR, 
as the excellent performance of the FRS for the forward $(p,d)$ spectroscopy 
has been demonstrated in the present experiment.
A large-acceptance 
detector, 
such as the WASA central detector \cite{WASA_CELSIUS, WASA_COSY}, 
will be additionally installed surrounding the reaction target. 
An experimental setup combining the FRS and the WASA systems
is illustrated in Fig.~\ref{fig_setup_semiexclusive}.
We will also consider possibilities of using
 other reaction channels such as the ($\pi,N$) reaction.

\section{Conclusion \label{s_conclusion}}
 
A missing-mass spectroscopy 
experiment of the $^{12}$C($p$,$d$) reaction was performed
at a proton energy of 2.5 GeV
aiming at the search for $\eta'$-mesic nuclei. 
The excitation-energy spectra
of $^{11}$C nuclei 
were successfully obtained around the $\eta'$-meson production threshold 
with high statistical sensitivity 
and sufficient experimental resolution. 
As no distinct peak structure has been 
observed in the excitation-energy spectra,
upper limits on the formation cross sections 
of the $\eta'$-mesic nuclei
have been determined. 
A comparison with theoretically-predicted formation spectra
sets a stringent constraint on the $\eta'$-nucleus potential.  

The present work has established 
the applicability of 
the missing-mass spectroscopy of the ($p$,$d$) reaction
for studying in-medium properties of the $\eta^\prime$ meson.
The application to other mesons will be considered.
The experimental search for $\eta^\prime$-mesic nuclei 
will further proceed with the semi-exclusive measurement 
by simultaneously detecting the decay particles.

\begin{acknowledgments} 
The authors would like to acknowledge the support from
the GSI staffs, 
particularly the accelerator group, 
the FRS engineering group, 
and the target laboratory.
We are grateful to staffs of Institut f\"ur Kernphysik, Forschungszentrum J\"ulich
for their support in a detector test at COSY. 
We also thank Dr.~M.~Tabata for specially developed aerogel.
Y.K.T.~acknowledges financial support
from Grant-in-Aid for JSPS Fellows (No.\ 258155) and 
H.F. from Kyoto University Young Scholars Overseas Visit Program.
This work is partly supported by a MEXT Grants-in-Aid for Scientific 
Research on Innovative Areas (Nos.\ JP24105705 and JP24105712),
JSPS Grants-in-Aid for Scientific Research (S) (No.\ JP23224008)
and for Young Scientists (A) (No.\ JP25707018),
by the National Natural Science Foundation of China (No.~11235002)
and by the Bundesministerium f\"ur Bildung und Forschung. 

\end{acknowledgments}

\end{document}